\documentclass[reprint,superscriptaddress,secnumarabic,floatfix,amsmath,amssymb, nobibnotes,aps,pre]{revtex4-2}
\usepackage{varioref}
\usepackage[dvipsnames]{xcolor}
\usepackage{hyperref}
\hypersetup{
linkcolor=BrickRed
,citecolor=Green
,filecolor=Mulberry
,urlcolor=NavyBlue
,menucolor=BrickRed
,runcolor=Mulberry
,linkbordercolor=BrickRed
,citebordercolor=Green
,filebordercolor=Mulberry
,urlbordercolor=NavyBlue
,menubordercolor=BrickRed
,runbordercolor=Mulberry
}
\usepackage[capitalize]{cleveref}

\newcommand{\mycomment}[1]{}
\usepackage{graphicx}
\usepackage{dcolumn}
\usepackage{bm}
\usepackage{color,soul}
\newcommand{\ro}  { \mathbf{r}}
\usepackage[mathlines]{lineno}
\usepackage{easyReview}
\setreviewson 

\begin{document}
\linespread{1.0}
\preprint{APS/123-QED}

	\title{Interplay of distinct modes of charge regulation on poly-acid ionization and conformation}
\author{Souradeep Ghosh}
\email{souradeep@wustl.edu}
	\affiliation{
Department of Biomedical Engineering and Center for Biomolecular Condensates, Washington University in St.~Louis, St.~Louis, MO, USA
}
\author{Aritra Chowdhury}
\affiliation{
Department of Biochemistry, University of Zurich, Zurich 8057, Switzerland}
\author{Dylan T. Tomares}
	\affiliation{
Department of Biomedical Engineering and Center for Biomolecular Condensates, Washington University in St.~Louis, St.~Louis, MO, USA
}
\author{Benjamin Schuler}
\affiliation{
Department of Biochemistry, University of Zurich, Zurich 8057, Switzerland}
\affiliation{Department of Physics, University of Zurich, Zurich 8057, Switzerland}
\author{Arindam Kundagrami}
\email{arindam@iiserkol.ac.in}
\affiliation{
Deparment of Physical Sciences and Centre for Advanced Functional Materials, Indian Institute of Science Education and Research Kolkata, Mohanpur 741246, India}
\author{Rohit V.~Pappu}
\email{pappu@wustl.edu}
\affiliation{
Department of Biomedical Engineering and Center for Biomolecular Condensates, Washington University in St.~Louis, St.~Louis, MO, USA
}
	\begin{abstract}
		\textbf{\abstractname:}
        We adapt the Edwards–Muthukumar theoretical framework for a single polymer chain to investigate the interplay between proton binding and counterion condensation for poly-acids.  We find that changes to pH enable non-monotonic transitions between anti- and conventional polyelectrolyte behaviors. In the former, the net charge and the overall dimensions increase with increasing salt concentration, while the converse is true for conventional polyelectrolytes. The polymeric nature and local solvent polarization drive significant pK\textsubscript{a} shifts when compared to the values of reference monoacids. These pK\textsubscript{a} shifts are enhanced in semi-flexible chains.
	\end{abstract}
	
	\maketitle
    
\section{Introduction.} 

Poly-acids are ubiquitous among biomacromolecules \cite{Tanford1957,Sigler1988, Pak2016, muthu2017, Wang2025,Ghasemi2021,MullerSpth2010,hofmann2019,Bigman2022,King2024}.  One can model poly-acids by treating them as poly-anions of fixed charge \cite{Ha1992,delacruz2003, muthu2004, arindam2010, king2015, Lin2016, Ghosh2023, Chowdhury2023, Phillips2024, Ghosh2025}, and accounting for charge renormalization via counterion condensation \cite{Manning1969,manning1978,Anderson1982}. This mode of charge regulation arises from territorial binding of solution ions \cite{manning1978}. An additional mode of charge regulation derives from the site-specific binding of protons \cite{Overman1994,Antosiewicz1994,Borkovec2001,Ullmann2003,Hass2015,Mugnai2021,Ghasemi2021,Hoover2024}. The linear charge density and polymer flexibility, combined with the pH and salt concentration in solution can engender site-specific, context-dependent shifts in the pK\textsubscript{a} values of ionizable groups within poly-acids \cite{Tanford1957,Ullmann2003,Ullmann-Bombarda2010,Ullmann-Klingen2006,Holm-Reinauer2024}. 

Many intrinsically disordered proteins / regions (IDPs / IDRs) are essentially poly-acids, wherein more than $50\%$ of the amino acid sequences are either aspartic (D) or glutamic (E) acid residues \cite{Bigman2022,Lee2022,King2024-nucleus,King2024,Ruff2025}. In two- and multi-phase systems such as nucleoli, the high internal concentrations of proteins with D/E-tracts sets up a pH gradient creating acidic dense phases that coexist with the nucleoplasm that has a pH of 7.2 \cite{King2024}. These gradients, which can be reconstituted in vitro, are shown to be passive and result from density inhomogeneities created by phase separation via complex coacervation \cite{King2024,King2024-nucleus}. Fossat et al., recently showed that alternating acid- and base-rich blocks generate a multiplicity of charge states. These charge states are all equivalently compatible with high $\alpha$-helical contents \cite{Fossat2023}. As a result, the systems show a duality of high charge state heterogeneity, as measured by proton potentiometry, and conformational homogeneity as revealed by spectroscopy and atomistic simulations \cite{Fossat2023}. Emerging proton and ion potentiometry measurements show that poly-acid IDPs feature confounding non-monotonic variations of net charge with pH and salt concentration, showing both conventional and anti-polyelectrolyte behaviors \cite{Tomares2025}.  

Explicit accounting of the multiplicity of poly-acid charge states \cite{Fossat2019,Mugnai2021} leads to the \textit{q}-canonical ensemble \cite{Fossat2021-BJ, Fossat2023}. Populations of the accessible charge states can be extracted from numerical analysis of proton titration curves that are measured at different salt concentrations \cite{Fitch2002,Laguecir2006,Fossat2021-BJ,Fossat2023,Tomares2025}. Generating the accessible conformational ensemble requires simulations that enable exchanges between all the relevant charge microstates \cite{Fossat2019,Fossat2023}. The numbers of charge microstates that need to be sampled can be prohibitively large for poly-acids. This makes it challenging for simulations based on either the \textit{q}-canonical \cite{Fossat2023} or constant pH ensembles \cite{Lee2004,Khandogin2005,Buslaev2022,MartinsdeOliveira2022}. Given these inherent difficulties in computations based on atomistic simulations, we pursue an analytically tractable model to understand how the interplay between the two modes of charge regulation of poly-acids enable an annealing of poly-acid charge. 

\begin{figure}[!htbp]%
	\centering
		 \includegraphics[width=0.5\linewidth]{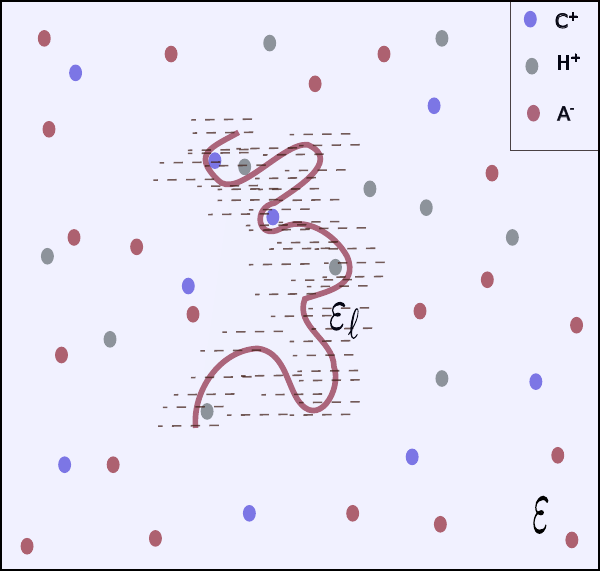}
		\caption{\textbf{Local-bulk partitioning of solution ions and protons around a poly-acid and setting up the separation ansatz:} Schematic representation of a single poly-acid chain immersed in a protic, salty solution. The polymer is modeled as a flexible random coil comprising $N$ Kuhn monomers. The local dielectric environment around the polymer differs from the bulk solvent, reflecting solvent polarization and preferential ion partitioning. Therefore, the local region is defined by a dielectric constant ($\varepsilon_l$) and it is demarcated from the bulk (dielectric constant $\varepsilon$) by the horizontal dashed lines. Mobile monovalent ions (C$^{+}$, A$^{-}$) and protons (H$^{+}$) are present in the local and bulk regions. Ions from the local region may be chosen for site-specific binding to the poly-acid (protonation) or for territorial binding around the poly-acid (ion condensation). Site-specific protonation and / or ion condensation will alter the charge on the poly-acid, and these alterations are mutually coupled to the intra-chain electrostatic interactions and to the interactions among ionic species. } \label{fig:schematic}
\end{figure}

\section{The Model}
A single poly-acid chain comprising $N$ Kuhn monomers is immersed in a protic solvent comprising a finite concentration ($c_s$) of monovalent salt ions. The solvent is treated as being a uniform dielectric material defined by a static dielectric constant $\varepsilon$ (\cref{fig:schematic}). The processes of interest are those that anneal the charge of the poly-acid via different interaction modes. The poly-acid in the dielectric continuum occupies a volume $\Omega$ at fixed pH and $c_s$. This volume can be demarcated into local and bulk regions \cite{Felitsky2004} (\cref{fig:schematic}). The canonical partition function of the system at temperature $T$ is written as $\mathcal{Z} = \mathcal{Z}_1 \mathcal{Z}_2 \mathcal{Z}_3$. Solution ions, including protons, can partition across the local and bulk regimes. Solution ions that partition into the local regime are either counterions or co-ions. Counterions can condense to alter the charge on the poly-acid. Conversely, if the chain is fully deprotonated, then co-ions will be excluded from the vicinity of the poly-acid. Protons that partition into the local regime can bind site-specifically to one of the $N_c$ sites on the poly-acid. The entropy of choosing from among $M_p$ protons and $M_s$ ions within the local layer will contribute to the term $\mathcal{Z}_1$.  The term $\mathcal{Z}_2$ quantifies the translational entropy of the hydrated ions within the volume $\Omega^{\prime} = \Omega\,(1 - \phi_\mathrm{ex})$. Here, $\phi_\mathrm{ex}$ accounts for the volume excluded by the polymer or ionic species. $\phi_\mathrm{ex}$ for the polymer is defined by its pervaded volume, and for solution ions and protons it is defined by the hydration volume \cite{Fossat2021}.  $\mathcal{Z}_3$ encompasses all other contributions: These include contributions from chain connectivity, excluded volume, screened intra-chain electrostatic interactions, the direct mean field free energy of proton binding to each of the ionizable sites, and the Coulombic mean field distance dependent free energy of ion pairing between solution ions and between solution ions and sites on the poly-acid. To avoid double counting, the Coulombic term excludes contributions from protons. 

The contour length of the chain is $L=N \ell$. The Helmholtz free energy ($F$) of the system is obtained using,


\begin{align}
	\label{partition-sum}
	e^{-\beta \mathcal{F}}=\mathcal{Z}_1 \mathcal{Z}_2 \exp \left(\frac{\Omega \kappa^3}{12 \pi}\right) \int \mathcal{D} \mathbf{R}(s)\exp \left[-\beta \left( \mathcal{H}+U_\mathrm{ad}\right)\right].
\end{align}

Here, $\beta=1 / k_B T$, $k_B$ is the Boltzmann constant, $T$ is the temperature, $\kappa$ is the screening parameter for electrostatic interactions, and $\mathbf{R}\left(s\right)$ is the position vector of the chain at the arc length variable $s$. The free energy $\mathcal{F}_\mathrm{pol}$ of the chain originating from $\mathcal{H}$ is written as

	\begin{align}\label{Partition_function}
		e^{-\beta \mathcal{F}_\mathrm{pol}}=\int\mathcal{D} \mathbf{R}\left(s\right) \exp (-\beta \mathcal{H}).
	\end{align}

	The Hamiltonian $\mathcal{H}$ encompasses contributions from chain connectivity ($\mathcal{H}_0$), short-range interactions among chain monomers ($\mathcal{H}_\mathrm{ex}$), and screened electrostatic interactions between uncompensated monomers ($\mathcal{H}_\mathrm{el}$). Accordingly,

\begin{align}
\mathcal{H}=\mathcal{H}_{0}+\mathcal{H}_\mathrm{ex}+\mathcal{H}_\mathrm{el},\label{eq:H}
\end{align}

with individual components defined as:

\begin{align}
\mathcal{H}_{0}&=\frac{3}{2 \ell^2}\int_{0}^{N} ds\left(\frac{\partial \mathbf{R}(s)}{\partial s}\right)^{2}\label{eq:H_0},\\[8pt]
\mathcal{H}_\mathrm{ex}&=\omega\ell^{3}\int_{0}^{N} ds \int_{0}^{N} ds^{\prime} \delta(\mathbf{R}(s)-\mathbf{R}(s^\prime))\label{eq:H_ex}, \text{and}\\[8pt]
\mathcal{H}_\mathrm{el}&=\frac{f^2\ell_{B}}{2}\int_{0}^{N} ds \int_{0}^{N} ds^\prime \frac{\exp(-\kappa|\mathbf{R}(s)-\mathbf{R}(s^\prime)|)}{|\mathbf{R}(s)-\mathbf{R}(s^\prime)|}\label{eq:H_el}.
\end{align}

Here, $\omega$ is the short-range excluded‐volume interaction parameter.  The inverse Debye length is $\tilde{\kappa}=\kappa\ell_{0}=({{{4 \pi  \tilde{\ell}_{B} \sum_{i} z_{i}^{2} n_{i} / \Omega}}})^{1/2}$, where  $n_i$ and $z_i$ represent the number and valency of the dissociated ions of the $i^{\text{th}}$ species; $\ell_0$ is the bond length and $\widetilde{\ell}_B( \equiv \ell_B/\ell_{0})$ is the dimensionless Bjerrum length ($l_B$).


 
\textit{The term $\mathcal{Z}_1$}: There are $N$ monomers and $N_c$ sites on the chain to which protons can bind or around which salt ions can condense. $\mathcal{Z}_1$ quantifies the combinatorial entropy of choosing among protons or counterions for binding to poly-acid sites.  $M_p$ protons and $M_s$ condensed counterions are drawn from the solution and therefore,
	
	\begin{align}
		\label{z1}
		\mathcal{Z}_1=\left[\frac{N_c !}{\left(N_c-M_p-M_s\right) ! M_{p} ! M_s !}\right].
	\end{align}
	
	\mycomment{The translational entropy of unadsorbed ions which are distributed in the volume $\Omega$, which we count as mobile ions, is computed by quantifying the number of arrangements of $N-M_p+n_{+}$ counterions, $n_{-}-M_{s}$ and $n_{-}$ monovalent co-ions. Therefore, the partition function related to the translational entropy inside the effective volume $\Omega$ is:
    
		\begin{align}
			\label{z2}
			\mathcal{Z}_2 &=\left[\frac{\left(\Omega^{\prime} / \ell^3\right)^{{\left(N-M_p\right)}+n_{+}+n_{-}}}{\left(N-M_p+n_{+}\right) !\left(n_{-}-M_{s}\right) !  n_{+} !}\right].
		\end{align}}

    \textit{The translational entropy $\mathcal{Z}_2$}: Next, we quantify the translational entropy of all the mobile ions in the volume $\Omega^{\prime}$. It is obtained by counting the arrangements of $n_1 = n_{+} - M_s$, $n_2 = N_c - M_p + n_{p}$, and $n_3 = n_{-}+n_{p}$ species where $n_{p}$ denotes the numbers of protons contributed by the protic solvent, and $n_{+}$ and $n_{-}$ are the numbers of cations and anions from the electrolyte (see \cref{appendix:free_energy_components} for details). The calculation of the translational entropy is based on previous approaches \cite{muthu2004,Kundagrami2008,Muthukumar2012,Ghosh2024}. $\mathcal{Z}_2$, which quantifies the translational entropy in the effective volume, is written as,
    
    \begin{equation}
    \label{z2}
     \mathcal{Z}_2 = \frac{\Omega^{\prime^{\sum_{\gamma=1}^{3} n_\gamma}}}{\prod_{\gamma=1}^{3} n_\gamma!}\,.
    \end{equation}

\textit{Components of $U_\mathrm{ad}$}: The term $U_\mathrm{ad}$ in Eq.~(\ref{partition-sum}) captures the contributions from site-specific binding of protons and territorial binding (condensation) of salt ions.  For the proton, we consider two processes. The mean-field free energy of protonating a single site is defined in terms of the ionization equilibrium of a monoacid.  We define $\tilde F_{H^+}$ as the dimensionless mean-field binding free energy of a proton to a monoacid that is referenced to the standard state. The activity of the protons is set by the pH. We use pH $\approx -\log_{10} ([\mathrm{H}^+] ~/ ~1 ~\mathrm{ M}),$ where $[\mathrm{H}^+]$ is the molar concentration of protons (in mol/L). Following Refs. \cite{Nozaki1967,muthu2017,Muthukumar2023-book}, we set $\tilde F_{H^+} = 2.303 \mathrm{pK_{int}}$. Here, pK\textsubscript{int} is the intrinsic pK\textsubscript{a} of the monoacid. Since a poly-acid is a concatenation of monoacids, the dimensionless mean-field free energy due to binding of $M_p$ protons is written as $-M_p\tilde F_{H^+} =-M_{p}(2.303\,\mathrm{pK_{int}})$. 

The second part of $U_\mathrm{ad}$ is the electrostatic mean-field interaction free energy between a negatively charged monomer and a cation drawn from the monovalent salt. For this, we use a distance-dependent ion-pairing term of the form,  $-e^2/(4 \pi \epsilon_0 \epsilon_l d)$. Here, $d$ denotes the dipole length of the ion-pair and $\epsilon_l$ is the local dielectric constant representing solvent polarization near ion-pairs. The pair energy is proportional to the product of the dielectric mismatch parameter $\delta =\epsilon/\epsilon_{l}$ and the dimensionless Bjerrum length $\widetilde{\ell}_B$, yielding a dimensionless electrostatic energy for the system (see \cref{appendix:binding_constant} for details). Taken together, the dimensionless mean field interaction free energy for ion adsorption becomes 
\begin{align}
\label{uad}
    \beta U_\mathrm{ad} = - M_s \delta \widetilde{\ell}_B/\widetilde{d} - M_{p}(2.303\mathrm{pK_{int}}).
\end{align}

Overall, the construction of $\mathcal{F}$ is based on a separation ansatz, whereby proton binding is purely short-range and is treated as being distinct from ion condensation. Charge regulation on poly-acid sites involves choosing a mode of regulation, which we capture via $\mathcal{Z}_1$. Site-specific binding of protons is captured using ionization equilibria of monoacids, and a mean-field approach of scaling via a multiplicative factor $M_p$ for poly-acids. Solution ions exhibit territorial binding \cite{Manning1969}, and this is treated using established approaches \cite{Muthukumar2023-book,Phillips2024,arindam2010,muthu2004,Ghosh2025}. The separation ansatz was mandated by emerging data from proton and ion potentiometry of ordered and disordered proteins \cite{Fossat2021,Tomares2025}, the preferential partitioning of protons into \cite{King2024,Hoffmann2025,Ausserwger2024}, the preferential exclusion of cations from \cite{Pant2025}, and differential partitioning of solution ions across coacervates \cite{Posey2024,Smokers2025}. Note that joint effects on intra-chain electrostatic interactions are modeled via the $\mathcal{H}_{el}$ term. 

The separation ansatz draws inspiration from computations based on the ABSINTH implicit solvent model and forcefield paradigm \cite{Vitalis2008,Choi2019,Fossat2019,Fossat2023}. In this approach, polypeptides are parsed into distinct, non-overlapping solvation groups. The chain as a whole is a union of solvation groups. For a fixed simulation temperature, each solvation group is assigned a reference free energy of solvation that is derived from experimental measurements for model compounds. In ABSINTH, the intrinsic free energy of solvation can be modulated by changing the extent of solvation, which is computed by conformation-specific solvent accessible volume fractions. In a similar vein, we use the intrinsic pK\textsubscript{a} of the monoacid to prescribe the free energy of binding of a proton to a single site. The variational calculation that we describe below, allows for changes to chain conformation and optimization of the total free energy. This generates the annealed charge state of the system by joint optimization of proton binding and counterion condensation. For each ionizable site $i$, the pK\textsubscript{a,$i$} is the pH at which the the probability of the site being protonated versus deprotonated is equivalent. Departures of  site-specific pK\textsubscript{a} values from those of the monoacid are predicted by the model to be pH- and $c_s$-specific, and these are responses that come from minimizing the total free energy.

\textit{Minimization of the free energy}: The total free energy (\cref{partition-sum}) for a single poly-acid chain is minimized with respect to $M_p$, $M_s$, and ${\ell}_1$, as outlined in \cref{eq:free_energy_sum_minimise}. It is valid for all degrees of ionization and temperatures providing $c_s\leq(8\pi\ell_B^3)^{-1}$. We use a variational approach to compute the free energy of the poly-acid in solution. For this, we introduce a trial Hamiltonian $\mathcal{H}_\mathrm{trial}$ (\cref{appendix:evaluation_of_F5}, Eq.\eqref{trial-Hamiltonian}) with the variational parameter $\ell_1$ representing an effective expansion factor of the chain. We then  compute the average of $(\mathcal{H}-\mathcal{H}_\mathrm{trial})$ and the interaction terms under the trial ensemble (Eqs.\eqref{F5-trial},~\eqref{intra-chain-H-ex}, \eqref{intra-chain-H-el}), considering a Gaussian distribution given by Eq.\eqref{rho-def}. Under the assumption of isotropic expansion or contraction, the average dimensionless radius of gyration of the chain is given by $R_g = \sqrt{N \ell_1 / 6}$~\cite{edwards1979,muthu1987,muthu2004,Muthukumar2023-book}. For further details, see \cref{appendix:evaluation_of_F5,appendix-section:rho-fourier,appendix-section:Hex,appendix:structure-factor}.

\section{Results}
We quantified the effects of pH and monovalent salt concentration on charge regulation of a model poly-acid.  We set the parameters based on the number of ionizable residues in the intrinsically disordered protein, prothymosin $\alpha$ \cite{MullerSpth2010,Ruggeri2017,Borgia2018,Chowdhury2023}. Here, $N = 52$ approximates the number of ionizable residues. $N_c =44$ refers to the number of acidic residues. All other parameters that were used in the calculations are listed in Table~\ref{tab:parameters} (\cref{appendix:parameters}).

\begin{figure}[!htbp]%
	\centering
		\includegraphics[width=1.0\linewidth]{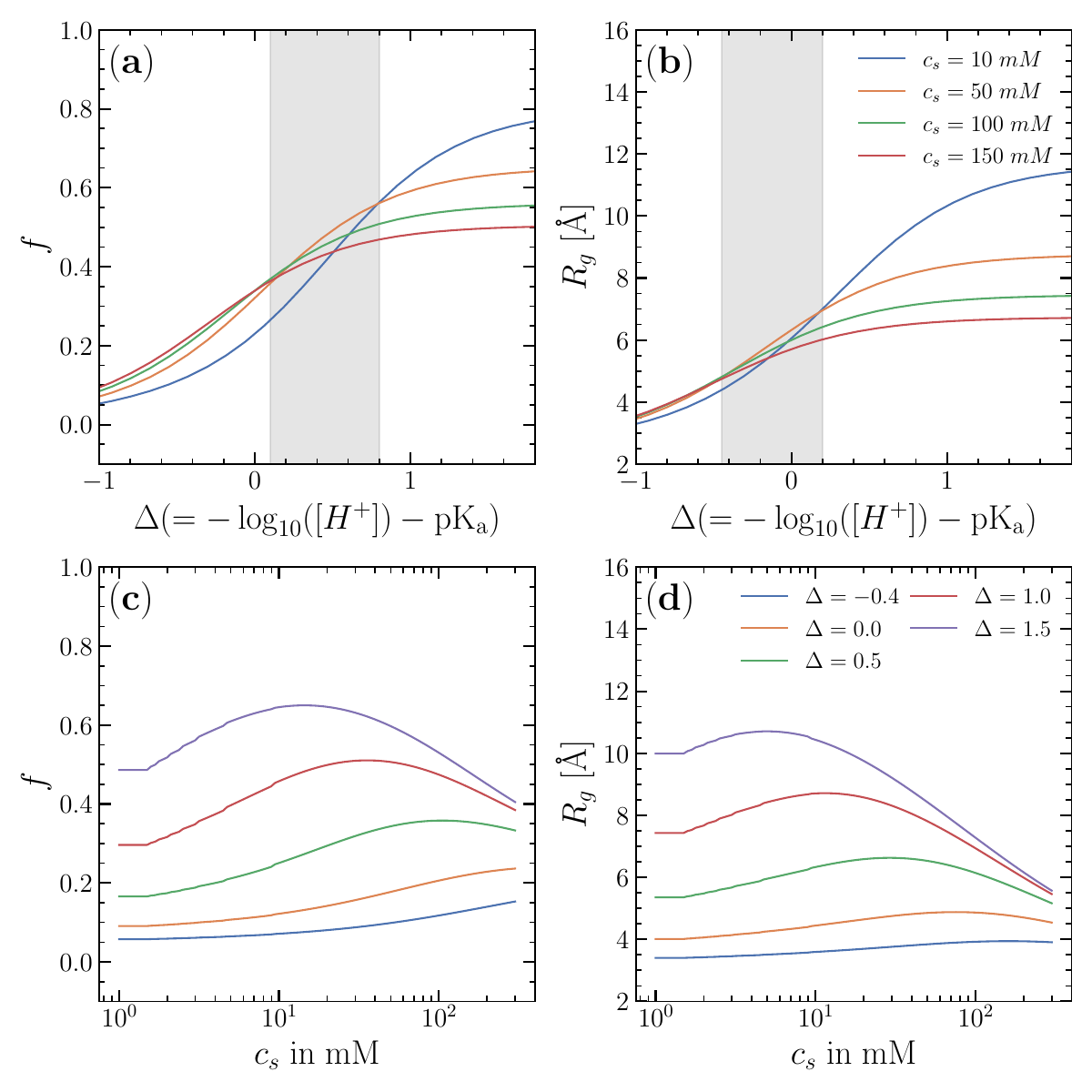}
		\caption{\textbf{Coupled titration-and-swelling responses of a poly-acid.}
(a) $f = N_c/N - M_{\mathrm p}/N - M_{s}/N$ and (b) $R_g$ as a function of $\Delta$ for different values of $c_s$. (c) $f$ and (d) $R_g$ versus salt concentration at fixed values of $\Delta$. The gray regions delineate different $\Delta$-windows in panels (a) versus (b).}\label{fig:fig-1} 
\end{figure}

We introduce a parameter $\Delta = -\log_{10}[H^+] -$ pK\textsubscript{a} \cite{Muthukumar2023-book}. At a given pH, $\Delta$ quantifies the pH differential referenced to the pK\textsubscript{a} of the monoacid. Positive values of $\Delta$ refer to pH values that are above the pK\textsubscript{a} of the monoacid, whereas negative $\Delta$ values represent the converse scenario. For fixed $c_s$, we quantified the extent of charge regulation in terms of the fraction of uncompensated acids along the poly-acid ($f = N_c/N - M_{\mathrm p}/N - M_{s}/N$). Low values of \textit{f} signify higher extents of charge regulation and vice versa. 

\textbf{Anti- and conventional polyelectrolyte behaviors}: Sigmoidal transitions were observed for \textit{f} (\cref{fig:fig-1}(a)) and $R_g$ as a function of $\Delta$ (\cref{fig:fig-1}(b)). When the pH is below the pK\textsubscript{a} of the monoacid (negative values of $\Delta$), \textit{f} is lowest at low salt and it increases monotonically with increasing $c_s$. However, when $\Delta$ is positive, we observe a non-monotonic variation of $f$ with $c_s$. This is made clear by analyzing the changes of \textit{f} and $R_g$ as a function of $c_s$ for fixed positive values of $\Delta$ (\cref{fig:fig-1}(c) and (d)). The non-monotonicity derives from there being distinct types of behaviors in the low versus high salt regimes. In the low-salt regime for positive $\Delta$, the poly-acid shows anti-polyelectrolyte behavior \cite{Jia2022,Liu2023}, whereby \textit{f} and $R_g$ increase with increasing $c_s$. Conventional polyelectrolyte behavior is recovered in the high-salt regime, whereby both \textit{f} and $R_g$ decrease with increasing $c_s$. For fixed $\Delta$, reentrant behavior sets in at higher $c_s$ for $f$ when compared to $R_g$. Thus $R_g$ is more sensitive to changes in salt concentration when compared to $f$. At intermediate values of $f$, screening of charge repulsions drives compaction because $f^2$ and $\kappa$ in \cref{eq:H_el} contribute jointly to lowering the charge repulsion. Thus, changes to charge and conformation can be decoupled from one another \cite{Fossat2023}.

\begin{figure}[!htbp]%
	\centering
		 \includegraphics[width=1.0\linewidth]{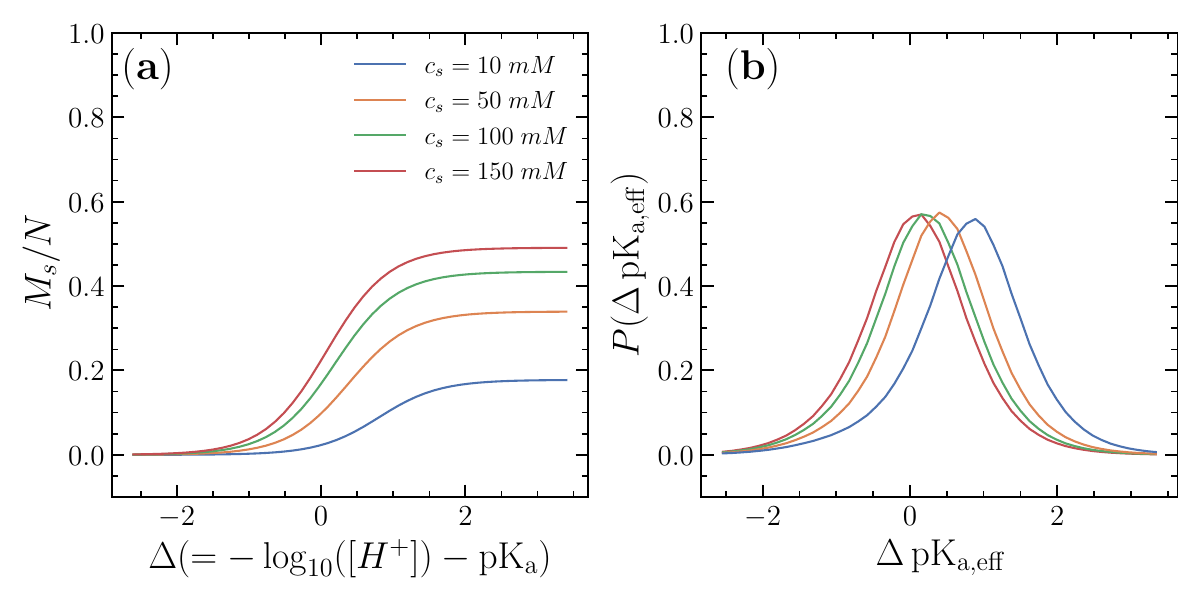}
		\caption{\textbf{Salt-dependent charge regulation of a poly-acid.}
(a) Fraction of condensed monovalent counterions, $M_s/N$, as a function of $\Delta$.
(b) Probability distribution of effective pK\textsubscript{a} shifts, $P$($\Delta$pK\textsubscript{a,eff}), at different salt concentrations. As pH increases, the polymer progressively deprotonates, exposing fixed negative charges. Higher salt concentrations enhance electrostatic screening thus promoting counterion condensation. This also causes a shift of the $P$($\Delta$pK\textsubscript{a,eff}) distribution so the mode approaches zero. The rise of $M_s/N$ in (a) reflects the interplay between proton binding / unbinding and ion condensation.} \label{fig:fig-2}
	\end{figure}

\textbf{pK\textsubscript{a} shifts enabled by chain connectivity:} The polymeric nature of a poly-acid leads to salt-dependent distributions of pK\textsubscript{a} values along the poly-acid (\cref{fig:fig-2}). The mode of the distribution of pK\textsubscript{a} values quantifies the magnitude of the shift of the average pK\textsubscript{a} vis-\'a-vis the monoacid. The mode shifts toward the monoacid pK\textsubscript{a} as $c_s$ increases. However, the distribution of the pK\textsubscript{a} values is never a delta function \cite{Mugnai2021,Fossat2021}. Instead, we observe a broad distribution of pK\textsubscript{a} values even for $c_s$ values of 150 mM.  

The binding capacity quantifies fluctuations in the number of ligands bound to a macromolecule, and is measured as a probability density function \cite{Di_Cera1993-kv}. In our case, the ligand is the proton, and the binding capacity of protons is manifest as a probability density distribution of pK\textsubscript{a} values. Proton binding to site $i$ is defined by a microscopic equilibrium constant $K_i$. The pH at which $K_i=1$ is the pK\textsubscript{a} value for site $i$ and it is designated as pK\textsubscript{a$i$}. This is used to compute the distribution of computed $\Delta$pK\textsubscript{a,eff} values, where $\Delta\mathrm{pK_{a,eff}}$ = (pK\textsubscript{a$i$}--pK\textsubscript{a}) for each of the $N_c$ sites designated as $i$.


\begin{figure}[!htbp]%
	\centering
		\includegraphics[width=1.0\linewidth]{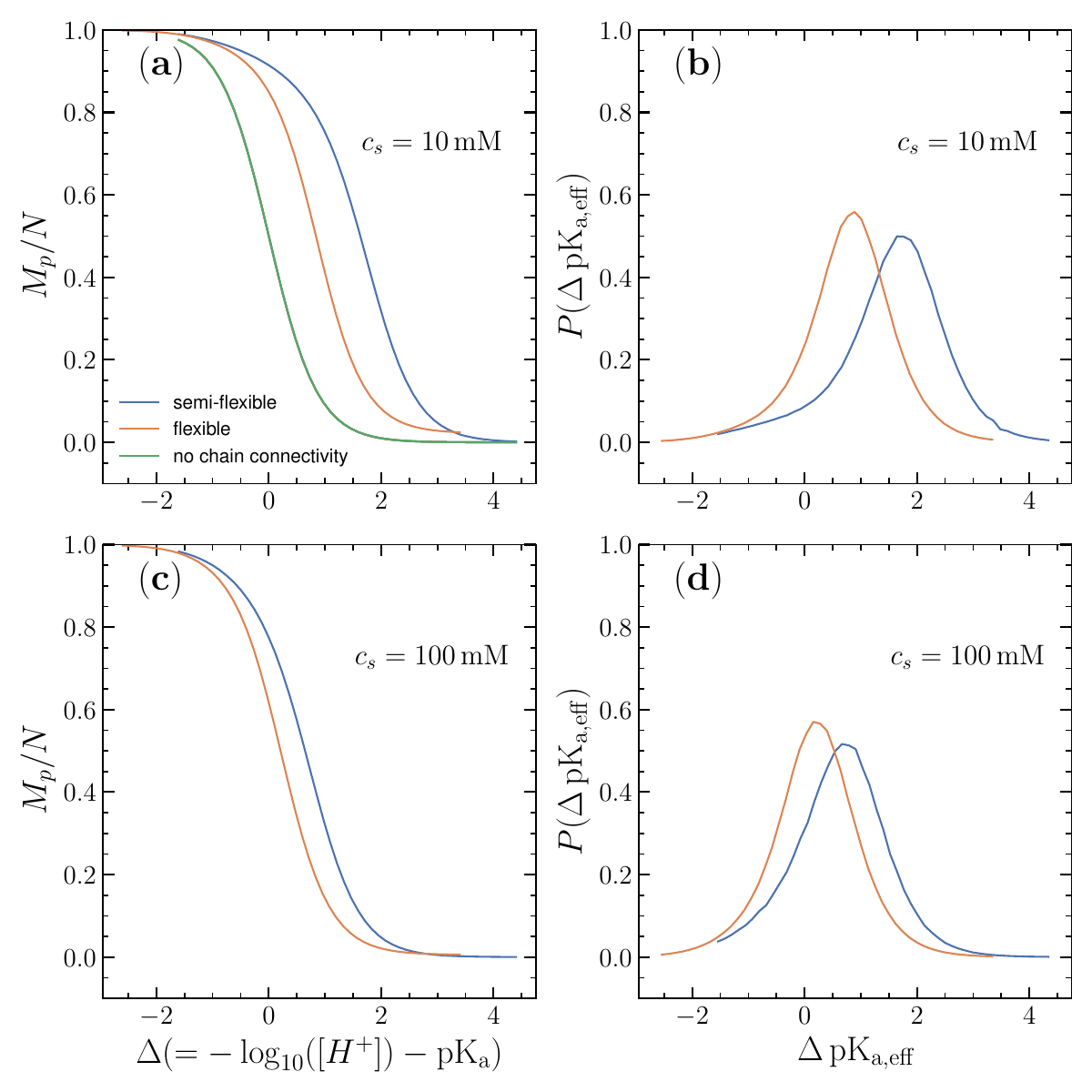}
		\caption{\textbf{Chain flexibility tunes charge-regulation in a poly-acid.} (a, c) Degree of protonation $M_p/N$ as a function of $\Delta$ for $c_s = 10$ mM (a) and $100$ mM (c). Results for semi-flexible and fully flexible chains are compared to expectations based on unshifted pk\textsubscript{a} values (no chain connectivity). (b,d) {$P$($\Delta$pK\textsubscript{a,eff})} distributions for flexible versus semi-flexible chains. With increasing salt concentration, both types of chains exhibit a monotonic, roughly logarithmic decrease of $\Delta$pK\textsubscript{a,eff}. However, the mode of the distribution approaches zero more slowly for the semi-flexible chain.} \label{fig:fig-3}
	\end{figure}

The polymeric nature of the poly-acid affects the set points of the rheostat that influence the pK\textsubscript{a} shifts. This is evident at low values of $c_s$ (\cref{fig:fig-3} (a) versus \cref{fig:fig-3}(c)). These observations are concordant with the work of Zhou who showed that a Gaussian chain model helps explain pK\textsubscript{a} shifts in denatured proteins \cite{Zhou2002,Zhou2003}. Diminished electrostatic screening amplifies intramolecular electrostatic repulsions, causing shifts of pK\textsubscript{a} values when compared to the monoacid. Enhanced screening at higher $c_s$ diminishes polymer-mediated electrostatic repulsions, thus attenuating shifts of mean pK\textsubscript{a} values.  

\textbf{Impact of chain flexibility:} Next, we assessed the impact of chain flexibility of poly-acids by comparing results for flexible versus semi-flexible chains \cite{Bhattacharjee1987, Ghosh2001} (see \cref{appendix:semi-flex} for details). The Kuhn length is invariant and semi-flexibility is introduced via local stiffness and bending rigidity. Semi-flexibility amplifies deviations from expectations based on monoacids (\cref{fig:fig-3} (a)-(d)). A stiffer chain forces neighboring ionizable groups to be persistently proximal to one another. Hence, the penalty for deprotonating an acidic group becomes larger than would be expected for flexible chains resulting in more pronounced pK\textsubscript{a} shifts (\cref{fig:fig-3} (b), (d)). 

For intrinsically disordered proteins, the persistence lengths are likely to span a range depending on the extent of linear clustering of acidic groups along the sequence \cite{Das2013, Pak2016,king2015,Lin2016,sing2020}, the type of acidic residue (Asp vs. Glu) \cite{Fossat2019,Choi2019-BASICS,Zeng2022-jz}, and the presence of and the contexts of proline versus glycine residues \cite{Cheng2010,Martin2016,Holla2024}. Therefore, changing the overall fraction of acidic residues, the linear clustering versus dispersion of these residues, the type of acidic residue, and / or the proline versus glycine content provides an encodable way of tuning the coupling between flexibility and charge regulation.  Sequence-encoded variations in chain flexibility can modulate the position and steepness of the protonation switch, thereby dictating how a poly-acid responds to local pH and ionic micro-environments. 

\begin{figure}[!htbp]%
	\centering
		\includegraphics[width=0.5\linewidth]{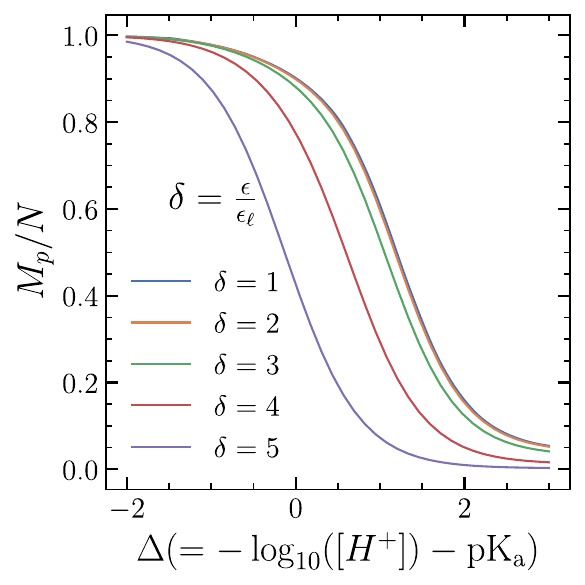}
		\caption{\textbf{Local dielectric inhomogeneity tunes charge-regulation in poly-acids.} Degree of protonation $M_p/N$ versus $\Delta$ for different values of $\delta$. Note that $\delta=1$ for $\epsilon=\epsilon_l$.} \label{fig:fig-5}
	\end{figure}

\textbf{Impact of solvent polarization:} Local dielectric inhomogeneities can arise from changes to the polarization of solvent molecules in the vicinity of the poly-acid \cite{Vitalis2009,Muthukumar2023-book}. We queried the effects of these inhomogeneities by varying the parameter $\delta = {\epsilon}/{\epsilon_l}$. Here, $\epsilon$ and $\epsilon_l$ are the bulk and local dielectric constants, respectively. If $\epsilon > \epsilon_l$, then solvent polarization in the vicinity of the poly-acid is different from the bulk and $\delta > 1$.  This mismatch between the local and bulk dielectric contributes to an enhancement of charge regulation, such that the extent of protonation of acidic groups shifts to larger, more positive values of $\Delta$ as compared to the case when $\delta=1$ (\cref{fig:fig-5}).  

\section{Summary}
We introduced an approach to model the interplay of proton binding, counterion condensation, dielectric inhomogeneities, and chain flexibility on the charge states and conformations of poly-acids. Our formulation represents a minimal yet analytically tractable approach for modeling and inferring the effects of charge regulation on poly-acid charge and conformation. Here, we extended the Edwards–Muthukumar variational framework to enable for the simultaneous treatment of site-specific proton binding and counterion condensation. The separation ansatz allows us to disentangle the two modes of charge regulation while consistently incorporating chain connectivity, excluded volume, screened electrostatics, and dielectric inhomogeneity. 

Application of the theory yields predictions of non-monotonic salt-dependent responses that include switching between conventional and anti-polyelectrolyte regimes, as well as broad, salt-dependent shifts of pK\textsubscript{a} values relative to that of monoacids. These results are consistent with emerging findings from potentiometric titrations \cite{Fossat2021-BJ,Tomares2025}, and measurements in two-phase systems \cite{King2024,Hoffmann2025}. Semi-flexible chains amplify the effects of charge regulation due to persistent electrostatic couplings along the backbone. 

Overall, our findings provide a statistical mechanical basis for interpreting potentiometric and spectroscopic measurements on intrinsically disordered proteins enriched in acidic residues, and highlight how sequence, flexibility, and local dielectric properties jointly govern the interplay of pH, salt, and conformation in disordered polymers. The approach can be extended to study charge regulation in two- and multiphase systems, where proton and ion gradients have been measured \cite{King2024,Posey2024,Hoffmann2025,Smokers2025,Ausserwger2024}, and different modes of charge regulation are likely to be significant.

\section*{Acknowledgments}
SG and RVP thank Martin Fossat, Gaurav Mitra, Soumik Mitra, Avnika Pant, Ammon Posey, and Kiersten Ruff for insightful discussions. RVP is grateful to Professor M. Muthukumar for pedagogical guidance. This work was supported by grants from the US National Science Foundation (MCB-2227268 to RVP), the Swiss National Science Foundation (to BS), and by IISER Kolkata, Ministry of Education, Government of India (to AK).
\bibliography{main}

\begin{thebibliography}{93}%
\makeatletter
\providecommand \@ifxundefined [1]{%
 \@ifx{#1\undefined}
}%
\providecommand \@ifnum [1]{%
 \ifnum #1\expandafter \@firstoftwo
 \else \expandafter \@secondoftwo
 \fi
}%
\providecommand \@ifx [1]{%
 \ifx #1\expandafter \@firstoftwo
 \else \expandafter \@secondoftwo
 \fi
}%
\providecommand \natexlab [1]{#1}%
\providecommand \enquote  [1]{``#1''}%
\providecommand \bibnamefont  [1]{#1}%
\providecommand \bibfnamefont [1]{#1}%
\providecommand \citenamefont [1]{#1}%
\providecommand \href@noop [0]{\@secondoftwo}%
\providecommand \href [0]{\begingroup \@sanitize@url \@href}%
\providecommand \@href[1]{\@@startlink{#1}\@@href}%
\providecommand \@@href[1]{\endgroup#1\@@endlink}%
\providecommand \@sanitize@url [0]{\catcode `\\12\catcode `\$12\catcode
  `\&12\catcode `\#12\catcode `\^12\catcode `\_12\catcode `\%12\relax}%
\providecommand \@@startlink[1]{}%
\providecommand \@@endlink[0]{}%
\providecommand \url  [0]{\begingroup\@sanitize@url \@url }%
\providecommand \@url [1]{\endgroup\@href {#1}{\urlprefix }}%
\providecommand \urlprefix  [0]{URL }%
\providecommand \Eprint [0]{\href }%
\providecommand \doibase [0]{https://doi.org/}%
\providecommand \selectlanguage [0]{\@gobble}%
\providecommand \bibinfo  [0]{\@secondoftwo}%
\providecommand \bibfield  [0]{\@secondoftwo}%
\providecommand \translation [1]{[#1]}%
\providecommand \BibitemOpen [0]{}%
\providecommand \bibitemStop [0]{}%
\providecommand \bibitemNoStop [0]{.\EOS\space}%
\providecommand \EOS [0]{\spacefactor3000\relax}%
\providecommand \BibitemShut  [1]{\csname bibitem#1\endcsname}%
\let\auto@bib@innerbib\@empty
\bibitem [{\citenamefont {Tanford}\ and\ \citenamefont
  {Kirkwood}(1957)}]{Tanford1957}%
  \BibitemOpen
  \bibfield  {author} {\bibinfo {author} {\bibfnamefont {C.}~\bibnamefont
  {Tanford}}\ and\ \bibinfo {author} {\bibfnamefont {J.~G.}\ \bibnamefont
  {Kirkwood}},\ }\bibfield  {title} {\bibinfo {title} {Theory of protein
  titration curves. i. general equations for impenetrable spheres},\ }\href
  {https://doi.org/10.1021/ja01577a001} {\bibfield  {journal} {\bibinfo
  {journal} {Journal of the American Chemical Society}\ }\textbf {\bibinfo
  {volume} {79}},\ \bibinfo {pages} {5333–5339} (\bibinfo {year}
  {1957})}\BibitemShut {NoStop}%
\bibitem [{\citenamefont {Sigler}(1988)}]{Sigler1988}%
  \BibitemOpen
  \bibfield  {author} {\bibinfo {author} {\bibfnamefont {P.~B.}\ \bibnamefont
  {Sigler}},\ }\bibfield  {title} {\bibinfo {title} {Acid blobs and negative
  noodles},\ }\href {https://doi.org/10.1038/333210a0} {\bibfield  {journal}
  {\bibinfo  {journal} {Nature}\ }\textbf {\bibinfo {volume} {333}},\ \bibinfo
  {pages} {210–212} (\bibinfo {year} {1988})}\BibitemShut {NoStop}%
\bibitem [{\citenamefont {Pak}\ \emph {et~al.}(2016)\citenamefont {Pak},
  \citenamefont {Kosno}, \citenamefont {Holehouse}, \citenamefont {Padrick},
  \citenamefont {Mittal}, \citenamefont {Ali}, \citenamefont {Yunus},
  \citenamefont {Liu}, \citenamefont {Pappu},\ and\ \citenamefont
  {Rosen}}]{Pak2016}%
  \BibitemOpen
  \bibfield  {author} {\bibinfo {author} {\bibfnamefont {C.~W.}\ \bibnamefont
  {Pak}}, \bibinfo {author} {\bibfnamefont {M.}~\bibnamefont {Kosno}}, \bibinfo
  {author} {\bibfnamefont {A.~S.}\ \bibnamefont {Holehouse}}, \bibinfo {author}
  {\bibfnamefont {S.~B.}\ \bibnamefont {Padrick}}, \bibinfo {author}
  {\bibfnamefont {A.}~\bibnamefont {Mittal}}, \bibinfo {author} {\bibfnamefont
  {R.}~\bibnamefont {Ali}}, \bibinfo {author} {\bibfnamefont {A.~A.}\
  \bibnamefont {Yunus}}, \bibinfo {author} {\bibfnamefont {D.~R.}\ \bibnamefont
  {Liu}}, \bibinfo {author} {\bibfnamefont {R.~V.}\ \bibnamefont {Pappu}},\
  and\ \bibinfo {author} {\bibfnamefont {M.~K.}\ \bibnamefont {Rosen}},\
  }\bibfield  {title} {\bibinfo {title} {Sequence determinants of intracellular
  phase separation by complex coacervation of a disordered protein},\ }\href
  {https://doi.org/10.1016/j.molcel.2016.05.042} {\bibfield  {journal}
  {\bibinfo  {journal} {Molecular Cell}\ }\textbf {\bibinfo {volume} {63}},\
  \bibinfo {pages} {72–85} (\bibinfo {year} {2016})}\BibitemShut {NoStop}%
\bibitem [{\citenamefont {Muthukumar}(2017)}]{muthu2017}%
  \BibitemOpen
  \bibfield  {author} {\bibinfo {author} {\bibfnamefont {M.}~\bibnamefont
  {Muthukumar}},\ }\bibfield  {title} {\bibinfo {title} {50th anniversary
  perspective: A perspective on polyelectrolyte solutions},\ }\href
  {https://doi.org/10.1021/acs.macromol.7b01929} {\bibfield  {journal}
  {\bibinfo  {journal} {Macromolecules}\ }\textbf {\bibinfo {volume} {50}},\
  \bibinfo {pages} {9528–9560} (\bibinfo {year} {2017})}\BibitemShut
  {NoStop}%
\bibitem [{\citenamefont {Wang}\ \emph {et~al.}(2025)\citenamefont {Wang},
  \citenamefont {Levy},\ and\ \citenamefont {Iwahara}}]{Wang2025}%
  \BibitemOpen
  \bibfield  {author} {\bibinfo {author} {\bibfnamefont {X.}~\bibnamefont
  {Wang}}, \bibinfo {author} {\bibfnamefont {Y.}~\bibnamefont {Levy}},\ and\
  \bibinfo {author} {\bibfnamefont {J.}~\bibnamefont {Iwahara}},\ }\bibfield
  {title} {\bibinfo {title} {Competition between nucleic acids and
  intrinsically disordered regions within proteins},\ }\href
  {https://doi.org/10.1021/acs.accounts.5c00261} {\bibfield  {journal}
  {\bibinfo  {journal} {Accounts of Chemical Research}\ }\textbf {\bibinfo
  {volume} {58}},\ \bibinfo {pages} {2415–2424} (\bibinfo {year}
  {2025})}\BibitemShut {NoStop}%
\bibitem [{\citenamefont {Ghasemi}\ and\ \citenamefont
  {Larson}(2021)}]{Ghasemi2021}%
  \BibitemOpen
  \bibfield  {author} {\bibinfo {author} {\bibfnamefont {M.}~\bibnamefont
  {Ghasemi}}\ and\ \bibinfo {author} {\bibfnamefont {R.~G.}\ \bibnamefont
  {Larson}},\ }\bibfield  {title} {\bibinfo {title} {Role of electrostatic
  interactions in charge regulation of weakly dissociating polyacids},\ }\href
  {https://doi.org/10.1016/j.progpolymsci.2020.101322} {\bibfield  {journal}
  {\bibinfo  {journal} {Progress in Polymer Science}\ }\textbf {\bibinfo
  {volume} {112}},\ \bibinfo {pages} {101322} (\bibinfo {year}
  {2021})}\BibitemShut {NoStop}%
\bibitem [{\citenamefont {M\"{u}ller-Sp\"{a}th}\ \emph
  {et~al.}(2010)\citenamefont {M\"{u}ller-Sp\"{a}th}, \citenamefont {Soranno},
  \citenamefont {Hirschfeld}, \citenamefont {Hofmann}, \citenamefont
  {R\"{u}egger}, \citenamefont {Reymond}, \citenamefont {Nettels},\ and\
  \citenamefont {Schuler}}]{MullerSpth2010}%
  \BibitemOpen
  \bibfield  {author} {\bibinfo {author} {\bibfnamefont {S.}~\bibnamefont
  {M\"{u}ller-Sp\"{a}th}}, \bibinfo {author} {\bibfnamefont {A.}~\bibnamefont
  {Soranno}}, \bibinfo {author} {\bibfnamefont {V.}~\bibnamefont {Hirschfeld}},
  \bibinfo {author} {\bibfnamefont {H.}~\bibnamefont {Hofmann}}, \bibinfo
  {author} {\bibfnamefont {S.}~\bibnamefont {R\"{u}egger}}, \bibinfo {author}
  {\bibfnamefont {L.}~\bibnamefont {Reymond}}, \bibinfo {author} {\bibfnamefont
  {D.}~\bibnamefont {Nettels}},\ and\ \bibinfo {author} {\bibfnamefont
  {B.}~\bibnamefont {Schuler}},\ }\bibfield  {title} {\bibinfo {title} {Charge
  interactions can dominate the dimensions of intrinsically disordered
  proteins},\ }\href {https://doi.org/10.1073/pnas.1001743107} {\bibfield
  {journal} {\bibinfo  {journal} {Proceedings of the National Academy of
  Sciences}\ }\textbf {\bibinfo {volume} {107}},\ \bibinfo {pages}
  {14609–14614} (\bibinfo {year} {2010})}\BibitemShut {NoStop}%
\bibitem [{\citenamefont {Vancraenenbroeck}\ \emph {et~al.}(2019)\citenamefont
  {Vancraenenbroeck}, \citenamefont {Harel}, \citenamefont {Zheng},\ and\
  \citenamefont {Hofmann}}]{hofmann2019}%
  \BibitemOpen
  \bibfield  {author} {\bibinfo {author} {\bibfnamefont {R.}~\bibnamefont
  {Vancraenenbroeck}}, \bibinfo {author} {\bibfnamefont {Y.~S.}\ \bibnamefont
  {Harel}}, \bibinfo {author} {\bibfnamefont {W.}~\bibnamefont {Zheng}},\ and\
  \bibinfo {author} {\bibfnamefont {H.}~\bibnamefont {Hofmann}},\ }\bibfield
  {title} {\bibinfo {title} {Polymer effects modulate binding affinities in
  disordered proteins},\ }\href {https://doi.org/10.1073/pnas.1904997116}
  {\bibfield  {journal} {\bibinfo  {journal} {Proceedings of the National
  Academy of Sciences}\ }\textbf {\bibinfo {volume} {116}},\ \bibinfo {pages}
  {19506} (\bibinfo {year} {2019})}\BibitemShut {NoStop}%
\bibitem [{\citenamefont {Bigman}\ \emph {et~al.}(2022)\citenamefont {Bigman},
  \citenamefont {Iwahara},\ and\ \citenamefont {Levy}}]{Bigman2022}%
  \BibitemOpen
  \bibfield  {author} {\bibinfo {author} {\bibfnamefont {L.~S.}\ \bibnamefont
  {Bigman}}, \bibinfo {author} {\bibfnamefont {J.}~\bibnamefont {Iwahara}},\
  and\ \bibinfo {author} {\bibfnamefont {Y.}~\bibnamefont {Levy}},\ }\bibfield
  {title} {\bibinfo {title} {Negatively charged disordered regions are
  prevalent and functionally important across proteomes},\ }\href
  {https://doi.org/10.1016/j.jmb.2022.167660} {\bibfield  {journal} {\bibinfo
  {journal} {Journal of Molecular Biology}\ }\textbf {\bibinfo {volume}
  {434}},\ \bibinfo {pages} {167660} (\bibinfo {year} {2022})}\BibitemShut
  {NoStop}%
\bibitem [{\citenamefont {King}\ \emph
  {et~al.}(2024{\natexlab{a}})\citenamefont {King}, \citenamefont {Ruff},
  \citenamefont {Lin}, \citenamefont {Pant}, \citenamefont {Farag},
  \citenamefont {Lalmansingh}, \citenamefont {Wu}, \citenamefont {Fossat},
  \citenamefont {Ouyang}, \citenamefont {Lew}, \citenamefont {Lundberg},
  \citenamefont {Vahey},\ and\ \citenamefont {Pappu}}]{King2024}%
  \BibitemOpen
  \bibfield  {author} {\bibinfo {author} {\bibfnamefont {M.~R.}\ \bibnamefont
  {King}}, \bibinfo {author} {\bibfnamefont {K.~M.}\ \bibnamefont {Ruff}},
  \bibinfo {author} {\bibfnamefont {A.~Z.}\ \bibnamefont {Lin}}, \bibinfo
  {author} {\bibfnamefont {A.}~\bibnamefont {Pant}}, \bibinfo {author}
  {\bibfnamefont {M.}~\bibnamefont {Farag}}, \bibinfo {author} {\bibfnamefont
  {J.~M.}\ \bibnamefont {Lalmansingh}}, \bibinfo {author} {\bibfnamefont
  {T.}~\bibnamefont {Wu}}, \bibinfo {author} {\bibfnamefont {M.~J.}\
  \bibnamefont {Fossat}}, \bibinfo {author} {\bibfnamefont {W.}~\bibnamefont
  {Ouyang}}, \bibinfo {author} {\bibfnamefont {M.~D.}\ \bibnamefont {Lew}},
  \bibinfo {author} {\bibfnamefont {E.}~\bibnamefont {Lundberg}}, \bibinfo
  {author} {\bibfnamefont {M.~D.}\ \bibnamefont {Vahey}},\ and\ \bibinfo
  {author} {\bibfnamefont {R.~V.}\ \bibnamefont {Pappu}},\ }\bibfield  {title}
  {\bibinfo {title} {Macromolecular condensation organizes nucleolar sub-phases
  to set up a ph gradient},\ }\href
  {https://doi.org/10.1016/j.cell.2024.02.029} {\bibfield  {journal} {\bibinfo
  {journal} {Cell}\ }\textbf {\bibinfo {volume} {187}},\ \bibinfo {pages}
  {1889} (\bibinfo {year} {2024}{\natexlab{a}})}\BibitemShut {NoStop}%
\bibitem [{\citenamefont {Ha}\ and\ \citenamefont {Thirumalai}(1992)}]{Ha1992}%
  \BibitemOpen
  \bibfield  {author} {\bibinfo {author} {\bibfnamefont {B.-Y.}\ \bibnamefont
  {Ha}}\ and\ \bibinfo {author} {\bibfnamefont {D.}~\bibnamefont
  {Thirumalai}},\ }\bibfield  {title} {\bibinfo {title} {Conformations of a
  polyelectrolyte chain},\ }\href {https://doi.org/10.1103/physreva.46.r3012}
  {\bibfield  {journal} {\bibinfo  {journal} {Physical Review A}\ }\textbf
  {\bibinfo {volume} {46}},\ \bibinfo {pages} {R3012–R3015} (\bibinfo {year}
  {1992})}\BibitemShut {NoStop}%
\bibitem [{\citenamefont {Ermoshkin}\ and\ \citenamefont {Olvera de~la
  Cruz}(2003)}]{delacruz2003}%
  \BibitemOpen
  \bibfield  {author} {\bibinfo {author} {\bibfnamefont {A.~V.}\ \bibnamefont
  {Ermoshkin}}\ and\ \bibinfo {author} {\bibfnamefont {M.}~\bibnamefont {Olvera
  de~la Cruz}},\ }\bibfield  {title} {\bibinfo {title} {A modified random phase
  approximation of polyelectrolyte solutions},\ }\href
  {https://doi.org/10.1021/ma034148p} {\bibfield  {journal} {\bibinfo
  {journal} {Macromolecules}\ }\textbf {\bibinfo {volume} {36}},\ \bibinfo
  {pages} {7824} (\bibinfo {year} {2003})},\ \Eprint
  {https://arxiv.org/abs/https://doi.org/10.1021/ma034148p}
  {https://doi.org/10.1021/ma034148p} \BibitemShut {NoStop}%
\bibitem [{\citenamefont {Muthukumar}(2004)}]{muthu2004}%
  \BibitemOpen
  \bibfield  {author} {\bibinfo {author} {\bibfnamefont {M.}~\bibnamefont
  {Muthukumar}},\ }\bibfield  {title} {\bibinfo {title} {Theory of counter-ion
  condensation on flexible polyelectrolytes: Adsorption mechanism},\ }\href
  {https://doi.org/10.1063/1.1701839} {\bibfield  {journal} {\bibinfo
  {journal} {JCP}\ }\textbf {\bibinfo {volume} {120}},\ \bibinfo {pages} {9343}
  (\bibinfo {year} {2004})},\ \Eprint
  {https://arxiv.org/abs/https://doi.org/10.1063/1.1701839}
  {https://doi.org/10.1063/1.1701839} \BibitemShut {NoStop}%
\bibitem [{\citenamefont {Kundagrami}\ and\ \citenamefont
  {Muthukumar}(2010)}]{arindam2010}%
  \BibitemOpen
  \bibfield  {author} {\bibinfo {author} {\bibfnamefont {A.}~\bibnamefont
  {Kundagrami}}\ and\ \bibinfo {author} {\bibfnamefont {M.}~\bibnamefont
  {Muthukumar}},\ }\bibfield  {title} {\bibinfo {title} {Effective charge and
  coil-globule transition of a polyelectrolyte chain},\ }\href
  {https://doi.org/10.1021/ma9020888} {\bibfield  {journal} {\bibinfo
  {journal} {Macromolecules}\ }\textbf {\bibinfo {volume} {43}},\ \bibinfo
  {pages} {2574} (\bibinfo {year} {2010})},\ \Eprint
  {https://arxiv.org/abs/https://doi.org/10.1021/ma9020888}
  {https://doi.org/10.1021/ma9020888} \BibitemShut {NoStop}%
\bibitem [{\citenamefont {Sawle}\ and\ \citenamefont {Ghosh}(2015)}]{king2015}%
  \BibitemOpen
  \bibfield  {author} {\bibinfo {author} {\bibfnamefont {L.}~\bibnamefont
  {Sawle}}\ and\ \bibinfo {author} {\bibfnamefont {K.}~\bibnamefont {Ghosh}},\
  }\bibfield  {title} {\bibinfo {title} {A theoretical method to compute
  sequence dependent configurational properties in charged polymers and
  proteins},\ }\bibfield  {journal} {\bibinfo  {journal} {The Journal of
  Chemical Physics}\ }\textbf {\bibinfo {volume} {143}},\ \href
  {https://doi.org/10.1063/1.4929391} {10.1063/1.4929391} (\bibinfo {year}
  {2015})\BibitemShut {NoStop}%
\bibitem [{\citenamefont {Lin}\ \emph {et~al.}(2016)\citenamefont {Lin},
  \citenamefont {Forman-Kay},\ and\ \citenamefont {Chan}}]{Lin2016}%
  \BibitemOpen
  \bibfield  {author} {\bibinfo {author} {\bibfnamefont {Y.-H.}\ \bibnamefont
  {Lin}}, \bibinfo {author} {\bibfnamefont {J.~D.}\ \bibnamefont
  {Forman-Kay}},\ and\ \bibinfo {author} {\bibfnamefont {H.~S.}\ \bibnamefont
  {Chan}},\ }\bibfield  {title} {\bibinfo {title} {Sequence-specific
  polyampholyte phase separation in membraneless organelles},\ }\bibfield
  {journal} {\bibinfo  {journal} {Physical Review Letters}\ }\textbf {\bibinfo
  {volume} {117}},\ \href {https://doi.org/10.1103/physrevlett.117.178101}
  {10.1103/physrevlett.117.178101} (\bibinfo {year} {2016})\BibitemShut
  {NoStop}%
\bibitem [{\citenamefont {Ghosh}\ \emph {et~al.}(2023)\citenamefont {Ghosh},
  \citenamefont {Mitra},\ and\ \citenamefont {Kundagrami}}]{Ghosh2023}%
  \BibitemOpen
  \bibfield  {author} {\bibinfo {author} {\bibfnamefont {S.}~\bibnamefont
  {Ghosh}}, \bibinfo {author} {\bibfnamefont {S.}~\bibnamefont {Mitra}},\ and\
  \bibinfo {author} {\bibfnamefont {A.}~\bibnamefont {Kundagrami}},\ }\bibfield
   {title} {\bibinfo {title} {Polymer complexation: Partially ionizable
  asymmetric polyelectrolytes},\ }\bibfield  {journal} {\bibinfo  {journal}
  {The Journal of Chemical Physics}\ }\textbf {\bibinfo {volume} {158}},\ \href
  {https://doi.org/10.1063/5.0147323} {10.1063/5.0147323} (\bibinfo {year}
  {2023})\BibitemShut {NoStop}%
\bibitem [{\citenamefont {Chowdhury}\ \emph {et~al.}(2023)\citenamefont
  {Chowdhury}, \citenamefont {Borgia}, \citenamefont {Ghosh}, \citenamefont
  {Sottini}, \citenamefont {Mitra}, \citenamefont {Eapen}, \citenamefont
  {Borgia}, \citenamefont {Yang}, \citenamefont {Galvanetto}, \citenamefont
  {Ivanovic}, \citenamefont {Lukijanczuk}, \citenamefont {Zhu}, \citenamefont
  {Nettels}, \citenamefont {Kundagrami},\ and\ \citenamefont
  {Schuler}}]{Chowdhury2023}%
  \BibitemOpen
  \bibfield  {author} {\bibinfo {author} {\bibfnamefont {A.}~\bibnamefont
  {Chowdhury}}, \bibinfo {author} {\bibfnamefont {A.}~\bibnamefont {Borgia}},
  \bibinfo {author} {\bibfnamefont {S.}~\bibnamefont {Ghosh}}, \bibinfo
  {author} {\bibfnamefont {A.}~\bibnamefont {Sottini}}, \bibinfo {author}
  {\bibfnamefont {S.}~\bibnamefont {Mitra}}, \bibinfo {author} {\bibfnamefont
  {R.~S.}\ \bibnamefont {Eapen}}, \bibinfo {author} {\bibfnamefont {M.~B.}\
  \bibnamefont {Borgia}}, \bibinfo {author} {\bibfnamefont {T.}~\bibnamefont
  {Yang}}, \bibinfo {author} {\bibfnamefont {N.}~\bibnamefont {Galvanetto}},
  \bibinfo {author} {\bibfnamefont {M.~T.}\ \bibnamefont {Ivanovic}}, \bibinfo
  {author} {\bibfnamefont {P.}~\bibnamefont {Lukijanczuk}}, \bibinfo {author}
  {\bibfnamefont {R.}~\bibnamefont {Zhu}}, \bibinfo {author} {\bibfnamefont
  {D.}~\bibnamefont {Nettels}}, \bibinfo {author} {\bibfnamefont
  {A.}~\bibnamefont {Kundagrami}},\ and\ \bibinfo {author} {\bibfnamefont
  {B.}~\bibnamefont {Schuler}},\ }\bibfield  {title} {\bibinfo {title} {Driving
  forces of the complex formation between highly charged disordered proteins},\
  }\bibfield  {journal} {\bibinfo  {journal} {Proceedings of the National
  Academy of Sciences}\ }\textbf {\bibinfo {volume} {120}},\ \href
  {https://doi.org/10.1073/pnas.2304036120} {10.1073/pnas.2304036120} (\bibinfo
  {year} {2023})\BibitemShut {NoStop}%
\bibitem [{\citenamefont {Phillips}\ \emph {et~al.}(2024)\citenamefont
  {Phillips}, \citenamefont {Muthukumar},\ and\ \citenamefont
  {Ghosh}}]{Phillips2024}%
  \BibitemOpen
  \bibfield  {author} {\bibinfo {author} {\bibfnamefont {M.}~\bibnamefont
  {Phillips}}, \bibinfo {author} {\bibfnamefont {M.}~\bibnamefont
  {Muthukumar}},\ and\ \bibinfo {author} {\bibfnamefont {K.}~\bibnamefont
  {Ghosh}},\ }\bibfield  {title} {\bibinfo {title} {Beyond monopole
  electrostatics in regulating conformations of intrinsically disordered
  proteins},\ }\bibfield  {journal} {\bibinfo  {journal} {PNAS Nexus}\ }\textbf
  {\bibinfo {volume} {3}},\ \href {https://doi.org/10.1093/pnasnexus/pgae367}
  {10.1093/pnasnexus/pgae367} (\bibinfo {year} {2024})\BibitemShut {NoStop}%
\bibitem [{\citenamefont {Ghosh}(2025)}]{Ghosh2025}%
  \BibitemOpen
  \bibfield  {author} {\bibinfo {author} {\bibfnamefont {S.}~\bibnamefont
  {Ghosh}},\ }\bibfield  {title} {\bibinfo {title} {Thermodynamic insights into
  polyelectrolyte complexation: A theoretical framework},\ }\bibfield
  {journal} {\bibinfo  {journal} {The Journal of Chemical Physics}\ }\textbf
  {\bibinfo {volume} {162}},\ \href {https://doi.org/10.1063/5.0250546}
  {10.1063/5.0250546} (\bibinfo {year} {2025})\BibitemShut {NoStop}%
\bibitem [{\citenamefont {Manning}(1969)}]{Manning1969}%
  \BibitemOpen
  \bibfield  {author} {\bibinfo {author} {\bibfnamefont {G.~S.}\ \bibnamefont
  {Manning}},\ }\bibfield  {title} {\bibinfo {title} {Limiting laws and
  counterion condensation in polyelectrolyte solutions i. colligative
  properties},\ }\href {https://doi.org/10.1063/1.1672157} {\bibfield
  {journal} {\bibinfo  {journal} {The Journal of Chemical Physics}\ }\textbf
  {\bibinfo {volume} {51}},\ \bibinfo {pages} {924–933} (\bibinfo {year}
  {1969})}\BibitemShut {NoStop}%
\bibitem [{\citenamefont {Manning}(1978)}]{manning1978}%
  \BibitemOpen
  \bibfield  {author} {\bibinfo {author} {\bibfnamefont {G.~S.}\ \bibnamefont
  {Manning}},\ }\bibfield  {title} {\bibinfo {title} {The molecular theory of
  polyelectrolyte solutions with applications to the electrostatic properties
  of polynucleotides},\ }\href {https://doi.org/10.1017/s0033583500002031}
  {\bibfield  {journal} {\bibinfo  {journal} {Quarterly Reviews of Biophysics}\
  }\textbf {\bibinfo {volume} {11}},\ \bibinfo {pages} {179} (\bibinfo {year}
  {1978})}\BibitemShut {NoStop}%
\bibitem [{\citenamefont {Anderson}\ and\ \citenamefont
  {Record}(1982)}]{Anderson1982}%
  \BibitemOpen
  \bibfield  {author} {\bibinfo {author} {\bibfnamefont {C.~F.}\ \bibnamefont
  {Anderson}}\ and\ \bibinfo {author} {\bibfnamefont {M.~T.}\ \bibnamefont
  {Record}},\ }\bibfield  {title} {\bibinfo {title} {Polyelectrolyte theories
  and their applications to dna},\ }\href
  {https://doi.org/10.1146/annurev.pc.33.100182.001203} {\bibfield  {journal}
  {\bibinfo  {journal} {Annual Review of Physical Chemistry}\ }\textbf
  {\bibinfo {volume} {33}},\ \bibinfo {pages} {191–222} (\bibinfo {year}
  {1982})}\BibitemShut {NoStop}%
\bibitem [{\citenamefont {Overman}\ and\ \citenamefont
  {Lohman}(1994)}]{Overman1994}%
  \BibitemOpen
  \bibfield  {author} {\bibinfo {author} {\bibfnamefont {L.~B.}\ \bibnamefont
  {Overman}}\ and\ \bibinfo {author} {\bibfnamefont {T.~M.}\ \bibnamefont
  {Lohman}},\ }\bibfield  {title} {\bibinfo {title} {Linkage of ph, anion and
  cation effects in protein-nucleic acid equilibria},\ }\href
  {https://doi.org/10.1006/jmbi.1994.1126} {\bibfield  {journal} {\bibinfo
  {journal} {Journal of Molecular Biology}\ }\textbf {\bibinfo {volume}
  {236}},\ \bibinfo {pages} {165–178} (\bibinfo {year} {1994})}\BibitemShut
  {NoStop}%
\bibitem [{\citenamefont {Antosiewicz}\ \emph {et~al.}(1994)\citenamefont
  {Antosiewicz}, \citenamefont {McCammon},\ and\ \citenamefont
  {Gilson}}]{Antosiewicz1994}%
  \BibitemOpen
  \bibfield  {author} {\bibinfo {author} {\bibfnamefont {J.}~\bibnamefont
  {Antosiewicz}}, \bibinfo {author} {\bibfnamefont {J.}~\bibnamefont
  {McCammon}},\ and\ \bibinfo {author} {\bibfnamefont {M.~K.}\ \bibnamefont
  {Gilson}},\ }\bibfield  {title} {\bibinfo {title} {Prediction of ph-dependent
  properties of proteins},\ }\href {https://doi.org/10.1006/jmbi.1994.1301}
  {\bibfield  {journal} {\bibinfo  {journal} {Journal of Molecular Biology}\
  }\textbf {\bibinfo {volume} {238}},\ \bibinfo {pages} {415–436} (\bibinfo
  {year} {1994})}\BibitemShut {NoStop}%
\bibitem [{\citenamefont {Borkovec}\ \emph {et~al.}(2001)\citenamefont
  {Borkovec}, \citenamefont {J\"{o}nsson},\ and\ \citenamefont
  {Koper}}]{Borkovec2001}%
  \BibitemOpen
  \bibfield  {author} {\bibinfo {author} {\bibfnamefont {M.}~\bibnamefont
  {Borkovec}}, \bibinfo {author} {\bibfnamefont {B.}~\bibnamefont
  {J\"{o}nsson}},\ and\ \bibinfo {author} {\bibfnamefont {G.~J.~M.}\
  \bibnamefont {Koper}},\ }\bibinfo {title} {Ionization processes and proton
  binding in polyprotic systems: Small molecules, proteins, interfaces, and
  polyelectrolytes},\ in\ \href {https://doi.org/10.1007/978-1-4615-1223-3_2}
  {\emph {\bibinfo {booktitle} {Surface and Colloid Science}}}\ (\bibinfo
  {publisher} {Springer US},\ \bibinfo {year} {2001})\ p.\ \bibinfo {pages}
  {99–339}\BibitemShut {NoStop}%
\bibitem [{\citenamefont {Ullmann}(2003)}]{Ullmann2003}%
  \BibitemOpen
  \bibfield  {author} {\bibinfo {author} {\bibfnamefont {G.~M.}\ \bibnamefont
  {Ullmann}},\ }\bibfield  {title} {\bibinfo {title} {Relations between
  protonation constants and titration curves in polyprotic acids: A critical
  view},\ }\href {https://doi.org/10.1021/jp026454v} {\bibfield  {journal}
  {\bibinfo  {journal} {The Journal of Physical Chemistry B}\ }\textbf
  {\bibinfo {volume} {107}},\ \bibinfo {pages} {1263–1271} (\bibinfo {year}
  {2003})}\BibitemShut {NoStop}%
\bibitem [{\citenamefont {Hass}\ and\ \citenamefont {Mulder}(2015)}]{Hass2015}%
  \BibitemOpen
  \bibfield  {author} {\bibinfo {author} {\bibfnamefont {M.~A.}\ \bibnamefont
  {Hass}}\ and\ \bibinfo {author} {\bibfnamefont {F.~A.}\ \bibnamefont
  {Mulder}},\ }\bibfield  {title} {\bibinfo {title} {Contemporary nmr studies
  of protein electrostatics},\ }\href
  {https://doi.org/10.1146/annurev-biophys-083012-130351} {\bibfield  {journal}
  {\bibinfo  {journal} {Annual Review of Biophysics}\ }\textbf {\bibinfo
  {volume} {44}},\ \bibinfo {pages} {53–75} (\bibinfo {year}
  {2015})}\BibitemShut {NoStop}%
\bibitem [{\citenamefont {Mugnai}\ and\ \citenamefont
  {Thirumalai}(2021)}]{Mugnai2021}%
  \BibitemOpen
  \bibfield  {author} {\bibinfo {author} {\bibfnamefont {M.~L.}\ \bibnamefont
  {Mugnai}}\ and\ \bibinfo {author} {\bibfnamefont {D.}~\bibnamefont
  {Thirumalai}},\ }\bibfield  {title} {\bibinfo {title} {Molecular transfer
  model for ph effects on intrinsically disordered proteins: Theory and
  applications},\ }\href {https://doi.org/10.1021/acs.jctc.0c01316} {\bibfield
  {journal} {\bibinfo  {journal} {Journal of Chemical Theory and Computation}\
  }\textbf {\bibinfo {volume} {17}},\ \bibinfo {pages} {1944–1954} (\bibinfo
  {year} {2021})}\BibitemShut {NoStop}%
\bibitem [{\citenamefont {Hoover}\ \emph {et~al.}(2024)\citenamefont {Hoover},
  \citenamefont {Margossian},\ and\ \citenamefont {Muthukumar}}]{Hoover2024}%
  \BibitemOpen
  \bibfield  {author} {\bibinfo {author} {\bibfnamefont {S.~C.}\ \bibnamefont
  {Hoover}}, \bibinfo {author} {\bibfnamefont {K.~O.}\ \bibnamefont
  {Margossian}},\ and\ \bibinfo {author} {\bibfnamefont {M.}~\bibnamefont
  {Muthukumar}},\ }\bibfield  {title} {\bibinfo {title} {Theory and
  quantitative assessment of ph-responsive polyzwitterion–polyelectrolyte
  complexation},\ }\href {https://doi.org/10.1039/d4sm00575a} {\bibfield
  {journal} {\bibinfo  {journal} {Soft Matter}\ }\textbf {\bibinfo {volume}
  {20}},\ \bibinfo {pages} {7199–7213} (\bibinfo {year} {2024})}\BibitemShut
  {NoStop}%
\bibitem [{\citenamefont {Bombarda}\ and\ \citenamefont
  {Ullmann}(2010)}]{Ullmann-Bombarda2010}%
  \BibitemOpen
  \bibfield  {author} {\bibinfo {author} {\bibfnamefont {E.}~\bibnamefont
  {Bombarda}}\ and\ \bibinfo {author} {\bibfnamefont {G.~M.}\ \bibnamefont
  {Ullmann}},\ }\bibfield  {title} {\bibinfo {title} {ph-dependent pka values
  in proteins—a theoretical analysis of protonation energies with practical
  consequences for enzymatic reactions},\ }\href
  {https://doi.org/10.1021/jp908926w} {\bibfield  {journal} {\bibinfo
  {journal} {The Journal of Physical Chemistry B}\ }\textbf {\bibinfo {volume}
  {114}},\ \bibinfo {pages} {1994–2003} (\bibinfo {year} {2010})}\BibitemShut
  {NoStop}%
\bibitem [{\citenamefont {Klingen}\ \emph {et~al.}(2006)\citenamefont
  {Klingen}, \citenamefont {Bombarda},\ and\ \citenamefont
  {Ullmann}}]{Ullmann-Klingen2006}%
  \BibitemOpen
  \bibfield  {author} {\bibinfo {author} {\bibfnamefont {A.~R.}\ \bibnamefont
  {Klingen}}, \bibinfo {author} {\bibfnamefont {E.}~\bibnamefont {Bombarda}},\
  and\ \bibinfo {author} {\bibfnamefont {G.~M.}\ \bibnamefont {Ullmann}},\
  }\bibfield  {title} {\bibinfo {title} {Theoretical investigation of the
  behavior of titratable groups in proteins},\ }\href
  {https://doi.org/10.1039/b515479k} {\bibfield  {journal} {\bibinfo  {journal}
  {Photochemical and Photobiological Sciences}\ }\textbf {\bibinfo {volume}
  {5}},\ \bibinfo {pages} {588–596} (\bibinfo {year} {2006})}\BibitemShut
  {NoStop}%
\bibitem [{\citenamefont {Reinauer}\ \emph {et~al.}(2024)\citenamefont
  {Reinauer}, \citenamefont {Kondrat},\ and\ \citenamefont
  {Holm}}]{Holm-Reinauer2024}%
  \BibitemOpen
  \bibfield  {author} {\bibinfo {author} {\bibfnamefont {A.}~\bibnamefont
  {Reinauer}}, \bibinfo {author} {\bibfnamefont {S.}~\bibnamefont {Kondrat}},\
  and\ \bibinfo {author} {\bibfnamefont {C.}~\bibnamefont {Holm}},\ }\bibfield
  {title} {\bibinfo {title} {Electrolytes in conducting nanopores: Revisiting
  constant charge and constant potential simulations},\ }\bibfield  {journal}
  {\bibinfo  {journal} {The Journal of Chemical Physics}\ }\textbf {\bibinfo
  {volume} {161}},\ \href {https://doi.org/10.1063/5.0226959}
  {10.1063/5.0226959} (\bibinfo {year} {2024})\BibitemShut {NoStop}%
\bibitem [{\citenamefont {Lee}\ \emph {et~al.}(2022)\citenamefont {Lee},
  \citenamefont {Jaberi-Lashkari},\ and\ \citenamefont {Calo}}]{Lee2022}%
  \BibitemOpen
  \bibfield  {author} {\bibinfo {author} {\bibfnamefont {B.}~\bibnamefont
  {Lee}}, \bibinfo {author} {\bibfnamefont {N.}~\bibnamefont
  {Jaberi-Lashkari}},\ and\ \bibinfo {author} {\bibfnamefont {E.}~\bibnamefont
  {Calo}},\ }\bibfield  {title} {\bibinfo {title} {A unified view of low
  complexity regions (lcrs) across species},\ }\bibfield  {journal} {\bibinfo
  {journal} {eLife}\ }\textbf {\bibinfo {volume} {11}},\ \href
  {https://doi.org/10.7554/elife.77058} {10.7554/elife.77058} (\bibinfo {year}
  {2022})\BibitemShut {NoStop}%
\bibitem [{\citenamefont {King}\ \emph
  {et~al.}(2024{\natexlab{b}})\citenamefont {King}, \citenamefont {Ruff},\ and\
  \citenamefont {and}}]{King2024-nucleus}%
  \BibitemOpen
  \bibfield  {author} {\bibinfo {author} {\bibfnamefont {M.~R.}\ \bibnamefont
  {King}}, \bibinfo {author} {\bibfnamefont {K.~M.}\ \bibnamefont {Ruff}},\
  and\ \bibinfo {author} {\bibfnamefont {R.~V.~P.}\ \bibnamefont {and}},\
  }\bibfield  {title} {\bibinfo {title} {Emergent microenvironments of
  nucleoli},\ }\href {https://doi.org/10.1080/19491034.2024.2319957} {\bibfield
   {journal} {\bibinfo  {journal} {Nucleus}\ }\textbf {\bibinfo {volume}
  {15}},\ \bibinfo {pages} {2319957} (\bibinfo {year} {2024}{\natexlab{b}})},\
  \bibinfo {note} {pMID: 38443761},\ \Eprint
  {https://arxiv.org/abs/https://doi.org/10.1080/19491034.2024.2319957}
  {https://doi.org/10.1080/19491034.2024.2319957} \BibitemShut {NoStop}%
\bibitem [{\citenamefont {Ruff}\ \emph {et~al.}(2025)\citenamefont {Ruff},
  \citenamefont {King}, \citenamefont {Ying}, \citenamefont {Liu},
  \citenamefont {Pant}, \citenamefont {Lieberman}, \citenamefont {Shinn},\ and\
  \citenamefont {Pappu}}]{Ruff2025}%
  \BibitemOpen
  \bibfield  {author} {\bibinfo {author} {\bibfnamefont {K.~M.}\ \bibnamefont
  {Ruff}}, \bibinfo {author} {\bibfnamefont {M.~R.}\ \bibnamefont {King}},
  \bibinfo {author} {\bibfnamefont {A.~W.}\ \bibnamefont {Ying}}, \bibinfo
  {author} {\bibfnamefont {V.}~\bibnamefont {Liu}}, \bibinfo {author}
  {\bibfnamefont {A.}~\bibnamefont {Pant}}, \bibinfo {author} {\bibfnamefont
  {W.~E.}\ \bibnamefont {Lieberman}}, \bibinfo {author} {\bibfnamefont {S.~X.
  K.~C.}\ \bibnamefont {Shinn}, \bibfnamefont {Min~Kyung}},\ and\ \bibinfo
  {author} {\bibfnamefont {R.~V.}\ \bibnamefont {Pappu}},\ }\bibfield  {title}
  {\bibinfo {title} {Molecular grammars of intrinsically disordered regions
  that span the human proteome},\ }\bibfield  {journal} {\bibinfo  {journal}
  {bioRxiv}\ }\href {https://doi.org/10.1101/2025.02.27.640591}
  {10.1101/2025.02.27.640591} (\bibinfo {year} {2025})\BibitemShut {NoStop}%
\bibitem [{\citenamefont {Fossat}\ \emph {et~al.}(2023)\citenamefont {Fossat},
  \citenamefont {Posey},\ and\ \citenamefont {Pappu}}]{Fossat2023}%
  \BibitemOpen
  \bibfield  {author} {\bibinfo {author} {\bibfnamefont {M.~J.}\ \bibnamefont
  {Fossat}}, \bibinfo {author} {\bibfnamefont {A.~E.}\ \bibnamefont {Posey}},\
  and\ \bibinfo {author} {\bibfnamefont {R.~V.}\ \bibnamefont {Pappu}},\
  }\bibfield  {title} {\bibinfo {title} {Uncovering the contributions of charge
  regulation to the stability of single alpha helices},\ }\bibfield  {journal}
  {\bibinfo  {journal} {ChemPhysChem}\ }\textbf {\bibinfo {volume} {24}},\
  \href {https://doi.org/10.1002/cphc.202200746} {10.1002/cphc.202200746}
  (\bibinfo {year} {2023})\BibitemShut {NoStop}%
\bibitem [{\citenamefont {Tomares}\ \emph {et~al.}(2025)\citenamefont
  {Tomares}, \citenamefont {Ghosh}, \citenamefont {Palomo}, \citenamefont
  {Posey}, \citenamefont {King}, \citenamefont {Gupta}, \citenamefont
  {Fossat},\ and\ \citenamefont {Pappu}}]{Tomares2025}%
  \BibitemOpen
  \bibfield  {author} {\bibinfo {author} {\bibfnamefont {D.}~\bibnamefont
  {Tomares}}, \bibinfo {author} {\bibfnamefont {S.}~\bibnamefont {Ghosh}},
  \bibinfo {author} {\bibfnamefont {D.}~\bibnamefont {Palomo}}, \bibinfo
  {author} {\bibfnamefont {A.~E.}\ \bibnamefont {Posey}}, \bibinfo {author}
  {\bibfnamefont {M.~R.}\ \bibnamefont {King}}, \bibinfo {author}
  {\bibfnamefont {N.}~\bibnamefont {Gupta}}, \bibinfo {author} {\bibfnamefont
  {M.~J.}\ \bibnamefont {Fossat}},\ and\ \bibinfo {author} {\bibfnamefont
  {R.~V.}\ \bibnamefont {Pappu}},\ }\bibfield  {title} {\bibinfo {title}
  {Bps2025 - charge regulation of intrinsically disordered proteins},\ }\href
  {https://doi.org/10.1016/j.bpj.2024.11.2899} {\bibfield  {journal} {\bibinfo
  {journal} {Biophysical Journal}\ }\textbf {\bibinfo {volume} {124}},\
  \bibinfo {pages} {556a} (\bibinfo {year} {2025})}\BibitemShut {NoStop}%
\bibitem [{\citenamefont {Fossat}\ and\ \citenamefont
  {Pappu}(2019)}]{Fossat2019}%
  \BibitemOpen
  \bibfield  {author} {\bibinfo {author} {\bibfnamefont {M.~J.}\ \bibnamefont
  {Fossat}}\ and\ \bibinfo {author} {\bibfnamefont {R.~V.}\ \bibnamefont
  {Pappu}},\ }\bibfield  {title} {\bibinfo {title} {q-canonical monte carlo
  sampling for modeling the linkage between charge regulation and
  conformational equilibria of peptides},\ }\href
  {https://doi.org/10.1021/acs.jpcb.9b05206} {\bibfield  {journal} {\bibinfo
  {journal} {The Journal of Physical Chemistry B}\ }\textbf {\bibinfo {volume}
  {123}},\ \bibinfo {pages} {6952–6967} (\bibinfo {year} {2019})}\BibitemShut
  {NoStop}%
\bibitem [{\citenamefont {Fossat}\ \emph
  {et~al.}(2021{\natexlab{a}})\citenamefont {Fossat}, \citenamefont {Posey},\
  and\ \citenamefont {Pappu}}]{Fossat2021-BJ}%
  \BibitemOpen
  \bibfield  {author} {\bibinfo {author} {\bibfnamefont {M.~J.}\ \bibnamefont
  {Fossat}}, \bibinfo {author} {\bibfnamefont {A.~E.}\ \bibnamefont {Posey}},\
  and\ \bibinfo {author} {\bibfnamefont {R.~V.}\ \bibnamefont {Pappu}},\
  }\bibfield  {title} {\bibinfo {title} {Quantifying charge state heterogeneity
  for proteins with multiple ionizable residues},\ }\href
  {https://doi.org/10.1016/j.bpj.2021.11.2886} {\bibfield  {journal} {\bibinfo
  {journal} {Biophysical Journal}\ }\textbf {\bibinfo {volume} {120}},\
  \bibinfo {pages} {5438–5453} (\bibinfo {year}
  {2021}{\natexlab{a}})}\BibitemShut {NoStop}%
\bibitem [{\citenamefont {Fitch}\ \emph {et~al.}(2002)\citenamefont {Fitch},
  \citenamefont {Karp}, \citenamefont {Lee}, \citenamefont {Stites},
  \citenamefont {Lattman},\ and\ \citenamefont {García-Moreno}}]{Fitch2002}%
  \BibitemOpen
  \bibfield  {author} {\bibinfo {author} {\bibfnamefont {C.~A.}\ \bibnamefont
  {Fitch}}, \bibinfo {author} {\bibfnamefont {D.~A.}\ \bibnamefont {Karp}},
  \bibinfo {author} {\bibfnamefont {K.~K.}\ \bibnamefont {Lee}}, \bibinfo
  {author} {\bibfnamefont {W.~E.}\ \bibnamefont {Stites}}, \bibinfo {author}
  {\bibfnamefont {E.~E.}\ \bibnamefont {Lattman}},\ and\ \bibinfo {author}
  {\bibfnamefont {E.~B.}\ \bibnamefont {García-Moreno}},\ }\bibfield  {title}
  {\bibinfo {title} {Experimental pka values of buried residues: Analysis with
  continuum methods and role of water penetration},\ }\href
  {https://doi.org/10.1016/s0006-3495(02)75670-1} {\bibfield  {journal}
  {\bibinfo  {journal} {Biophysical Journal}\ }\textbf {\bibinfo {volume}
  {82}},\ \bibinfo {pages} {3289–3304} (\bibinfo {year} {2002})}\BibitemShut
  {NoStop}%
\bibitem [{\citenamefont {Laguecir}\ \emph {et~al.}(2006)\citenamefont
  {Laguecir}, \citenamefont {Ulrich}, \citenamefont {Labille}, \citenamefont
  {Fatin-Rouge}, \citenamefont {Stoll},\ and\ \citenamefont
  {Buffle}}]{Laguecir2006}%
  \BibitemOpen
  \bibfield  {author} {\bibinfo {author} {\bibfnamefont {A.}~\bibnamefont
  {Laguecir}}, \bibinfo {author} {\bibfnamefont {S.}~\bibnamefont {Ulrich}},
  \bibinfo {author} {\bibfnamefont {J.}~\bibnamefont {Labille}}, \bibinfo
  {author} {\bibfnamefont {N.}~\bibnamefont {Fatin-Rouge}}, \bibinfo {author}
  {\bibfnamefont {S.}~\bibnamefont {Stoll}},\ and\ \bibinfo {author}
  {\bibfnamefont {J.}~\bibnamefont {Buffle}},\ }\bibfield  {title} {\bibinfo
  {title} {Size and ph effect on electrical and conformational behavior of
  poly(acrylic acid): Simulation and experiment},\ }\href
  {https://doi.org/10.1016/j.eurpolymj.2005.11.023} {\bibfield  {journal}
  {\bibinfo  {journal} {European Polymer Journal}\ }\textbf {\bibinfo {volume}
  {42}},\ \bibinfo {pages} {1135–1144} (\bibinfo {year} {2006})}\BibitemShut
  {NoStop}%
\bibitem [{\citenamefont {Lee}\ \emph {et~al.}(2004)\citenamefont {Lee},
  \citenamefont {Salsbury},\ and\ \citenamefont {Brooks}}]{Lee2004}%
  \BibitemOpen
  \bibfield  {author} {\bibinfo {author} {\bibfnamefont {M.~S.}\ \bibnamefont
  {Lee}}, \bibinfo {author} {\bibfnamefont {F.~R.}\ \bibnamefont {Salsbury}},\
  and\ \bibinfo {author} {\bibfnamefont {C.~L.}\ \bibnamefont {Brooks}},\
  }\bibfield  {title} {\bibinfo {title} {Constant‐ph molecular dynamics using
  continuous titration coordinates},\ }\href
  {https://doi.org/10.1002/prot.20128} {\bibfield  {journal} {\bibinfo
  {journal} {Proteins: Structure, Function, and Bioinformatics}\ }\textbf
  {\bibinfo {volume} {56}},\ \bibinfo {pages} {738–752} (\bibinfo {year}
  {2004})}\BibitemShut {NoStop}%
\bibitem [{\citenamefont {Khandogin}\ and\ \citenamefont
  {Brooks}(2005)}]{Khandogin2005}%
  \BibitemOpen
  \bibfield  {author} {\bibinfo {author} {\bibfnamefont {J.}~\bibnamefont
  {Khandogin}}\ and\ \bibinfo {author} {\bibfnamefont {C.~L.}\ \bibnamefont
  {Brooks}},\ }\bibfield  {title} {\bibinfo {title} {Constant ph molecular
  dynamics with proton tautomerism},\ }\href
  {https://doi.org/10.1529/biophysj.105.061341} {\bibfield  {journal} {\bibinfo
   {journal} {Biophysical Journal}\ }\textbf {\bibinfo {volume} {89}},\
  \bibinfo {pages} {141–157} (\bibinfo {year} {2005})}\BibitemShut {NoStop}%
\bibitem [{\citenamefont {Buslaev}\ \emph {et~al.}(2022)\citenamefont
  {Buslaev}, \citenamefont {Aho}, \citenamefont {Jansen}, \citenamefont
  {Bauer}, \citenamefont {Hess},\ and\ \citenamefont {Groenhof}}]{Buslaev2022}%
  \BibitemOpen
  \bibfield  {author} {\bibinfo {author} {\bibfnamefont {P.}~\bibnamefont
  {Buslaev}}, \bibinfo {author} {\bibfnamefont {N.}~\bibnamefont {Aho}},
  \bibinfo {author} {\bibfnamefont {A.}~\bibnamefont {Jansen}}, \bibinfo
  {author} {\bibfnamefont {P.}~\bibnamefont {Bauer}}, \bibinfo {author}
  {\bibfnamefont {B.}~\bibnamefont {Hess}},\ and\ \bibinfo {author}
  {\bibfnamefont {G.}~\bibnamefont {Groenhof}},\ }\bibfield  {title} {\bibinfo
  {title} {Best practices in constant ph md simulations: Accuracy and
  sampling},\ }\href {https://doi.org/10.1021/acs.jctc.2c00517} {\bibfield
  {journal} {\bibinfo  {journal} {Journal of Chemical Theory and Computation}\
  }\textbf {\bibinfo {volume} {18}},\ \bibinfo {pages} {6134–6147} (\bibinfo
  {year} {2022})}\BibitemShut {NoStop}%
\bibitem [{\citenamefont {Martins~de Oliveira}\ \emph
  {et~al.}(2022)\citenamefont {Martins~de Oliveira}, \citenamefont {Liu},\ and\
  \citenamefont {Shen}}]{MartinsdeOliveira2022}%
  \BibitemOpen
  \bibfield  {author} {\bibinfo {author} {\bibfnamefont {V.}~\bibnamefont
  {Martins~de Oliveira}}, \bibinfo {author} {\bibfnamefont {R.}~\bibnamefont
  {Liu}},\ and\ \bibinfo {author} {\bibfnamefont {J.}~\bibnamefont {Shen}},\
  }\bibfield  {title} {\bibinfo {title} {Constant ph molecular dynamics
  simulations: Current status and recent applications},\ }\href
  {https://doi.org/10.1016/j.sbi.2022.102498} {\bibfield  {journal} {\bibinfo
  {journal} {Current Opinion in Structural Biology}\ }\textbf {\bibinfo
  {volume} {77}},\ \bibinfo {pages} {102498} (\bibinfo {year}
  {2022})}\BibitemShut {NoStop}%
\bibitem [{\citenamefont {Felitsky}\ and\ \citenamefont
  {Record}(2004)}]{Felitsky2004}%
  \BibitemOpen
  \bibfield  {author} {\bibinfo {author} {\bibfnamefont {D.~J.}\ \bibnamefont
  {Felitsky}}\ and\ \bibinfo {author} {\bibfnamefont {M.~T.}\ \bibnamefont
  {Record}},\ }\bibfield  {title} {\bibinfo {title} {Application of the
  local-bulk partitioning and competitive binding models to interpret
  preferential interactions of glycine betaine and urea with protein surface},\
  }\href {https://doi.org/10.1021/bi049862t} {\bibfield  {journal} {\bibinfo
  {journal} {Biochemistry}\ }\textbf {\bibinfo {volume} {43}},\ \bibinfo
  {pages} {9276–9288} (\bibinfo {year} {2004})}\BibitemShut {NoStop}%
\bibitem [{\citenamefont {Fossat}\ \emph
  {et~al.}(2021{\natexlab{b}})\citenamefont {Fossat}, \citenamefont {Zeng},\
  and\ \citenamefont {Pappu}}]{Fossat2021}%
  \BibitemOpen
  \bibfield  {author} {\bibinfo {author} {\bibfnamefont {M.~J.}\ \bibnamefont
  {Fossat}}, \bibinfo {author} {\bibfnamefont {X.}~\bibnamefont {Zeng}},\ and\
  \bibinfo {author} {\bibfnamefont {R.~V.}\ \bibnamefont {Pappu}},\ }\bibfield
  {title} {\bibinfo {title} {Uncovering differences in hydration free energies
  and structures for model compound mimics of charged side chains of amino
  acids},\ }\href {https://doi.org/10.1021/acs.jpcb.1c01073} {\bibfield
  {journal} {\bibinfo  {journal} {The Journal of Physical Chemistry B}\
  }\textbf {\bibinfo {volume} {125}},\ \bibinfo {pages} {4148–4161} (\bibinfo
  {year} {2021}{\natexlab{b}})}\BibitemShut {NoStop}%
\bibitem [{\citenamefont {Kundagrami}\ and\ \citenamefont
  {Muthukumar}(2008)}]{Kundagrami2008}%
  \BibitemOpen
  \bibfield  {author} {\bibinfo {author} {\bibfnamefont {A.}~\bibnamefont
  {Kundagrami}}\ and\ \bibinfo {author} {\bibfnamefont {M.}~\bibnamefont
  {Muthukumar}},\ }\bibfield  {title} {\bibinfo {title} {Theory of competitive
  counterion adsorption on flexible polyelectrolytes: Divalent salts},\
  }\bibfield  {journal} {\bibinfo  {journal} {The Journal of Chemical Physics}\
  }\textbf {\bibinfo {volume} {128}},\ \href
  {https://doi.org/10.1063/1.2940199} {10.1063/1.2940199} (\bibinfo {year}
  {2008})\BibitemShut {NoStop}%
\bibitem [{\citenamefont {Muthukumar}(2012)}]{Muthukumar2012}%
  \BibitemOpen
  \bibfield  {author} {\bibinfo {author} {\bibfnamefont {M.}~\bibnamefont
  {Muthukumar}},\ }\bibfield  {title} {\bibinfo {title} {Counterion adsorption
  theory of dilute polyelectrolyte solutions: Apparent molecular weight, second
  virial coefficient, and intermolecular structure factor},\ }\href
  {https://doi.org/10.1063/1.4736545} {\bibfield  {journal} {\bibinfo
  {journal} {The Journal of Chemical Physics}\ }\textbf {\bibinfo {volume}
  {137}},\ \bibinfo {pages} {034902} (\bibinfo {year} {2012})}\BibitemShut
  {NoStop}%
\bibitem [{\citenamefont {Ghosh}\ and\ \citenamefont
  {Kundagrami}(2024)}]{Ghosh2024}%
  \BibitemOpen
  \bibfield  {author} {\bibinfo {author} {\bibfnamefont {S.}~\bibnamefont
  {Ghosh}}\ and\ \bibinfo {author} {\bibfnamefont {A.}~\bibnamefont
  {Kundagrami}},\ }\bibfield  {title} {\bibinfo {title} {Effect of counterion
  size on polyelectrolyte conformations and thermodynamics},\ }\bibfield
  {journal} {\bibinfo  {journal} {The Journal of Chemical Physics}\ }\textbf
  {\bibinfo {volume} {160}},\ \href {https://doi.org/10.1063/5.0178233}
  {10.1063/5.0178233} (\bibinfo {year} {2024})\BibitemShut {NoStop}%
\bibitem [{\citenamefont {Nozaki}\ and\ \citenamefont
  {Tanford}(1967)}]{Nozaki1967}%
  \BibitemOpen
  \bibfield  {author} {\bibinfo {author} {\bibfnamefont {Y.}~\bibnamefont
  {Nozaki}}\ and\ \bibinfo {author} {\bibfnamefont {C.}~\bibnamefont
  {Tanford}},\ }\bibfield  {title} {\bibinfo {title} {Examination of titration
  behavior},\ }\href {https://doi.org/10.1016/s0076-6879(67)11088-4} {\bibfield
   {journal} {\bibinfo  {journal} {Methods in Enzymology}\ }\textbf {\bibinfo
  {volume} {11}},\ \bibinfo {pages} {715–734} (\bibinfo {year}
  {1967})}\BibitemShut {NoStop}%
\bibitem [{\citenamefont {Muthukumar}(2023)}]{Muthukumar2023-book}%
  \BibitemOpen
  \bibfield  {author} {\bibinfo {author} {\bibfnamefont {M.}~\bibnamefont
  {Muthukumar}},\ }\href {https://doi.org/10.1017/9781139046749} {\emph
  {\bibinfo {title} {Physics of Charged Macromolecules: Synthetic and
  Biological Systems}}}\ (\bibinfo  {publisher} {Cambridge University Press},\
  \bibinfo {year} {2023})\BibitemShut {NoStop}%
\bibitem [{\citenamefont {Hoffmann}\ \emph {et~al.}(2025)\citenamefont
  {Hoffmann}, \citenamefont {Ruff}, \citenamefont {Edu}, \citenamefont {Shinn},
  \citenamefont {Tromm}, \citenamefont {King}, \citenamefont {Pant},
  \citenamefont {Ausserw\"{o}ger}, \citenamefont {Morgan}, \citenamefont
  {Knowles}, \citenamefont {Pappu},\ and\ \citenamefont
  {Milovanovic}}]{Hoffmann2025}%
  \BibitemOpen
  \bibfield  {author} {\bibinfo {author} {\bibfnamefont {C.}~\bibnamefont
  {Hoffmann}}, \bibinfo {author} {\bibfnamefont {K.~M.}\ \bibnamefont {Ruff}},
  \bibinfo {author} {\bibfnamefont {I.~A.}\ \bibnamefont {Edu}}, \bibinfo
  {author} {\bibfnamefont {M.~K.}\ \bibnamefont {Shinn}}, \bibinfo {author}
  {\bibfnamefont {J.~V.}\ \bibnamefont {Tromm}}, \bibinfo {author}
  {\bibfnamefont {M.~R.}\ \bibnamefont {King}}, \bibinfo {author}
  {\bibfnamefont {A.}~\bibnamefont {Pant}}, \bibinfo {author} {\bibfnamefont
  {H.}~\bibnamefont {Ausserw\"{o}ger}}, \bibinfo {author} {\bibfnamefont
  {J.~R.}\ \bibnamefont {Morgan}}, \bibinfo {author} {\bibfnamefont {T.~P.}\
  \bibnamefont {Knowles}}, \bibinfo {author} {\bibfnamefont {R.~V.}\
  \bibnamefont {Pappu}},\ and\ \bibinfo {author} {\bibfnamefont
  {D.}~\bibnamefont {Milovanovic}},\ }\bibfield  {title} {\bibinfo {title}
  {Synapsin condensation is governed by sequence-encoded molecular grammars},\
  }\href {https://doi.org/10.1016/j.jmb.2025.168987} {\bibfield  {journal}
  {\bibinfo  {journal} {Journal of Molecular Biology}\ }\textbf {\bibinfo
  {volume} {437}},\ \bibinfo {pages} {168987} (\bibinfo {year}
  {2025})}\BibitemShut {NoStop}%
\bibitem [{\citenamefont {Ausserw\"{o}ger}\ \emph {et~al.}(2024)\citenamefont
  {Ausserw\"{o}ger}, \citenamefont {Scrutton}, \citenamefont {Fischer},
  \citenamefont {Sneideris}, \citenamefont {Qian}, \citenamefont
  {de~Csilléry}, \citenamefont {Baronaite}, \citenamefont {Saar},
  \citenamefont {Białek}, \citenamefont {Oeller}, \citenamefont {Krainer},
  \citenamefont {Franzmann}, \citenamefont {Wittmann}, \citenamefont
  {Iglesias-Artola}, \citenamefont {Invernizzi}, \citenamefont {Hyman},
  \citenamefont {Alberti}, \citenamefont {Lorenzen},\ and\ \citenamefont
  {Knowles}}]{Ausserwger2024}%
  \BibitemOpen
  \bibfield  {author} {\bibinfo {author} {\bibfnamefont {H.}~\bibnamefont
  {Ausserw\"{o}ger}}, \bibinfo {author} {\bibfnamefont {R.}~\bibnamefont
  {Scrutton}}, \bibinfo {author} {\bibfnamefont {C.~M.}\ \bibnamefont
  {Fischer}}, \bibinfo {author} {\bibfnamefont {T.}~\bibnamefont {Sneideris}},
  \bibinfo {author} {\bibfnamefont {D.}~\bibnamefont {Qian}}, \bibinfo {author}
  {\bibfnamefont {E.}~\bibnamefont {de~Csilléry}}, \bibinfo {author}
  {\bibfnamefont {I.}~\bibnamefont {Baronaite}}, \bibinfo {author}
  {\bibfnamefont {K.~L.}\ \bibnamefont {Saar}}, \bibinfo {author}
  {\bibfnamefont {A.~Z.}\ \bibnamefont {Białek}}, \bibinfo {author}
  {\bibfnamefont {M.}~\bibnamefont {Oeller}}, \bibinfo {author} {\bibfnamefont
  {G.}~\bibnamefont {Krainer}}, \bibinfo {author} {\bibfnamefont {T.~M.}\
  \bibnamefont {Franzmann}}, \bibinfo {author} {\bibfnamefont {S.}~\bibnamefont
  {Wittmann}}, \bibinfo {author} {\bibfnamefont {J.~M.}\ \bibnamefont
  {Iglesias-Artola}}, \bibinfo {author} {\bibfnamefont {G.}~\bibnamefont
  {Invernizzi}}, \bibinfo {author} {\bibfnamefont {A.~A.}\ \bibnamefont
  {Hyman}}, \bibinfo {author} {\bibfnamefont {S.}~\bibnamefont {Alberti}},
  \bibinfo {author} {\bibfnamefont {N.}~\bibnamefont {Lorenzen}},\ and\
  \bibinfo {author} {\bibfnamefont {T.~P.~J.}\ \bibnamefont {Knowles}},\
  }\bibfield  {title} {\bibinfo {title} {Biomolecular condensates sustain ph
  gradients at equilibrium through charge neutralisation}\ }\href
  {https://doi.org/10.1101/2024.05.23.595321} {10.1101/2024.05.23.595321}
  (\bibinfo {year} {2024})\BibitemShut {NoStop}%
\bibitem [{\citenamefont {Pant}\ \emph {et~al.}(2025)\citenamefont {Pant},
  \citenamefont {Tomares}, \citenamefont {King}, \citenamefont {Dai},\ and\
  \citenamefont {Pappu}}]{Pant2025}%
  \BibitemOpen
  \bibfield  {author} {\bibinfo {author} {\bibfnamefont {A.}~\bibnamefont
  {Pant}}, \bibinfo {author} {\bibfnamefont {D.}~\bibnamefont {Tomares}},
  \bibinfo {author} {\bibfnamefont {M.~R.}\ \bibnamefont {King}}, \bibinfo
  {author} {\bibfnamefont {Y.}~\bibnamefont {Dai}},\ and\ \bibinfo {author}
  {\bibfnamefont {R.~V.}\ \bibnamefont {Pappu}},\ }\bibfield  {title} {\bibinfo
  {title} {Bps2025 - composition-dependent interphase electrochemical
  potentials of biomolecular condensates and their functional consequences},\
  }\href {https://doi.org/10.1016/j.bpj.2024.11.1294} {\bibfield  {journal}
  {\bibinfo  {journal} {Biophysical Journal}\ }\textbf {\bibinfo {volume}
  {124}},\ \bibinfo {pages} {237a} (\bibinfo {year} {2025})}\BibitemShut
  {NoStop}%
\bibitem [{\citenamefont {Posey}\ \emph {et~al.}(2024)\citenamefont {Posey},
  \citenamefont {Bremer}, \citenamefont {Erkamp}, \citenamefont {Pant},
  \citenamefont {Knowles}, \citenamefont {Dai}, \citenamefont {Mittag},\ and\
  \citenamefont {Pappu}}]{Posey2024}%
  \BibitemOpen
  \bibfield  {author} {\bibinfo {author} {\bibfnamefont {A.~E.}\ \bibnamefont
  {Posey}}, \bibinfo {author} {\bibfnamefont {A.}~\bibnamefont {Bremer}},
  \bibinfo {author} {\bibfnamefont {N.~A.}\ \bibnamefont {Erkamp}}, \bibinfo
  {author} {\bibfnamefont {A.}~\bibnamefont {Pant}}, \bibinfo {author}
  {\bibfnamefont {T.~P.~J.}\ \bibnamefont {Knowles}}, \bibinfo {author}
  {\bibfnamefont {Y.}~\bibnamefont {Dai}}, \bibinfo {author} {\bibfnamefont
  {T.}~\bibnamefont {Mittag}},\ and\ \bibinfo {author} {\bibfnamefont {R.~V.}\
  \bibnamefont {Pappu}},\ }\bibfield  {title} {\bibinfo {title} {Biomolecular
  condensates are characterized by interphase electric potentials},\ }\href
  {https://doi.org/10.1021/jacs.4c08946} {\bibfield  {journal} {\bibinfo
  {journal} {Journal of the American Chemical Society}\ }\textbf {\bibinfo
  {volume} {146}},\ \bibinfo {pages} {28268–28281} (\bibinfo {year}
  {2024})}\BibitemShut {NoStop}%
\bibitem [{\citenamefont {Smokers}\ \emph {et~al.}(2025)\citenamefont
  {Smokers}, \citenamefont {Lavagna}, \citenamefont {Freire}, \citenamefont
  {Paloni}, \citenamefont {Voets}, \citenamefont {Barducci}, \citenamefont
  {White}, \citenamefont {Khajehpour},\ and\ \citenamefont
  {Spruijt}}]{Smokers2025}%
  \BibitemOpen
  \bibfield  {author} {\bibinfo {author} {\bibfnamefont {I.~B.~A.}\
  \bibnamefont {Smokers}}, \bibinfo {author} {\bibfnamefont {E.}~\bibnamefont
  {Lavagna}}, \bibinfo {author} {\bibfnamefont {R.~V.~M.}\ \bibnamefont
  {Freire}}, \bibinfo {author} {\bibfnamefont {M.}~\bibnamefont {Paloni}},
  \bibinfo {author} {\bibfnamefont {I.~K.}\ \bibnamefont {Voets}}, \bibinfo
  {author} {\bibfnamefont {A.}~\bibnamefont {Barducci}}, \bibinfo {author}
  {\bibfnamefont {P.~B.}\ \bibnamefont {White}}, \bibinfo {author}
  {\bibfnamefont {M.}~\bibnamefont {Khajehpour}},\ and\ \bibinfo {author}
  {\bibfnamefont {E.}~\bibnamefont {Spruijt}},\ }\bibfield  {title} {\bibinfo
  {title} {Selective ion binding and uptake shape the microenvironment of
  biomolecular condensates},\ }\href {https://doi.org/10.1021/jacs.5c07295}
  {\bibfield  {journal} {\bibinfo  {journal} {Journal of the American Chemical
  Society}\ }\textbf {\bibinfo {volume} {147}},\ \bibinfo {pages}
  {25692–25704} (\bibinfo {year} {2025})}\BibitemShut {NoStop}%
\bibitem [{\citenamefont {Vitalis}\ and\ \citenamefont
  {Pappu}(2008{\natexlab{a}})}]{Vitalis2008}%
  \BibitemOpen
  \bibfield  {author} {\bibinfo {author} {\bibfnamefont {A.}~\bibnamefont
  {Vitalis}}\ and\ \bibinfo {author} {\bibfnamefont {R.~V.}\ \bibnamefont
  {Pappu}},\ }\bibfield  {title} {\bibinfo {title} {Absinth: A new continuum
  solvation model for simulations of polypeptides in aqueous solutions},\
  }\href {https://doi.org/10.1002/jcc.21005} {\bibfield  {journal} {\bibinfo
  {journal} {Journal of Computational Chemistry}\ }\textbf {\bibinfo {volume}
  {30}},\ \bibinfo {pages} {673–699} (\bibinfo {year}
  {2008}{\natexlab{a}})}\BibitemShut {NoStop}%
\bibitem [{\citenamefont {Choi}\ and\ \citenamefont {Pappu}(2019)}]{Choi2019}%
  \BibitemOpen
  \bibfield  {author} {\bibinfo {author} {\bibfnamefont {J.-M.}\ \bibnamefont
  {Choi}}\ and\ \bibinfo {author} {\bibfnamefont {R.~V.}\ \bibnamefont
  {Pappu}},\ }\bibfield  {title} {\bibinfo {title} {Improvements to the absinth
  force field for proteins based on experimentally derived amino acid specific
  backbone conformational statistics},\ }\href
  {https://doi.org/10.1021/acs.jctc.8b00573} {\bibfield  {journal} {\bibinfo
  {journal} {Journal of Chemical Theory and Computation}\ }\textbf {\bibinfo
  {volume} {15}},\ \bibinfo {pages} {1367–1382} (\bibinfo {year}
  {2019})}\BibitemShut {NoStop}%
\bibitem [{\citenamefont {Edwards}\ and\ \citenamefont
  {Singh}(1979)}]{edwards1979}%
  \BibitemOpen
  \bibfield  {author} {\bibinfo {author} {\bibfnamefont {S.~F.}\ \bibnamefont
  {Edwards}}\ and\ \bibinfo {author} {\bibfnamefont {P.}~\bibnamefont
  {Singh}},\ }\bibfield  {title} {\bibinfo {title} {Size of a polymer molecule
  in solution. part 1. {E}xcluded {v}olume problem},\ }\href@noop {} {\bibfield
   {journal} {\bibinfo  {journal} {Journal of the Chemical Society, Faraday
  Transactions 2: Molecular and Chemical Physics}\ }\textbf {\bibinfo {volume}
  {75}},\ \bibinfo {pages} {1001} (\bibinfo {year} {1979})}\BibitemShut
  {NoStop}%
\bibitem [{\citenamefont {Muthukumar}(1987)}]{muthu1987}%
  \BibitemOpen
  \bibfield  {author} {\bibinfo {author} {\bibfnamefont {M.}~\bibnamefont
  {Muthukumar}},\ }\bibfield  {title} {\bibinfo {title} {Adsorption of a
  polyelectrolyte chain to a charged surface},\ }\href
  {https://doi.org/10.1063/1.452763} {\bibfield  {journal} {\bibinfo  {journal}
  {JCP}\ }\textbf {\bibinfo {volume} {86}},\ \bibinfo {pages} {7230} (\bibinfo
  {year} {1987})}\BibitemShut {NoStop}%
\bibitem [{\citenamefont {Ruggeri}\ \emph {et~al.}(2017)\citenamefont
  {Ruggeri}, \citenamefont {Zosel}, \citenamefont {Mutter}, \citenamefont
  {Różycka}, \citenamefont {Wojtas}, \citenamefont {Ożyhar}, \citenamefont
  {Schuler},\ and\ \citenamefont {Krishnan}}]{Ruggeri2017}%
  \BibitemOpen
  \bibfield  {author} {\bibinfo {author} {\bibfnamefont {F.}~\bibnamefont
  {Ruggeri}}, \bibinfo {author} {\bibfnamefont {F.}~\bibnamefont {Zosel}},
  \bibinfo {author} {\bibfnamefont {N.}~\bibnamefont {Mutter}}, \bibinfo
  {author} {\bibfnamefont {M.}~\bibnamefont {Różycka}}, \bibinfo {author}
  {\bibfnamefont {M.}~\bibnamefont {Wojtas}}, \bibinfo {author} {\bibfnamefont
  {A.}~\bibnamefont {Ożyhar}}, \bibinfo {author} {\bibfnamefont
  {B.}~\bibnamefont {Schuler}},\ and\ \bibinfo {author} {\bibfnamefont
  {M.}~\bibnamefont {Krishnan}},\ }\bibfield  {title} {\bibinfo {title}
  {Single-molecule electrometry},\ }\href
  {https://doi.org/10.1038/nnano.2017.26} {\bibfield  {journal} {\bibinfo
  {journal} {Nature Nanotechnology}\ }\textbf {\bibinfo {volume} {12}},\
  \bibinfo {pages} {488–495} (\bibinfo {year} {2017})}\BibitemShut {NoStop}%
\bibitem [{\citenamefont {Borgia}\ \emph {et~al.}(2018)\citenamefont {Borgia},
  \citenamefont {Borgia}, \citenamefont {Bugge}, \citenamefont {Kissling},
  \citenamefont {Heidarsson}, \citenamefont {Fernandes}, \citenamefont
  {Sottini}, \citenamefont {Soranno}, \citenamefont {Buholzer}, \citenamefont
  {Nettels}, \citenamefont {Kragelund}, \citenamefont {Best},\ and\
  \citenamefont {Schuler}}]{Borgia2018}%
  \BibitemOpen
  \bibfield  {author} {\bibinfo {author} {\bibfnamefont {A.}~\bibnamefont
  {Borgia}}, \bibinfo {author} {\bibfnamefont {M.~B.}\ \bibnamefont {Borgia}},
  \bibinfo {author} {\bibfnamefont {K.}~\bibnamefont {Bugge}}, \bibinfo
  {author} {\bibfnamefont {V.~M.}\ \bibnamefont {Kissling}}, \bibinfo {author}
  {\bibfnamefont {P.~O.}\ \bibnamefont {Heidarsson}}, \bibinfo {author}
  {\bibfnamefont {C.~B.}\ \bibnamefont {Fernandes}}, \bibinfo {author}
  {\bibfnamefont {A.}~\bibnamefont {Sottini}}, \bibinfo {author} {\bibfnamefont
  {A.}~\bibnamefont {Soranno}}, \bibinfo {author} {\bibfnamefont {K.~J.}\
  \bibnamefont {Buholzer}}, \bibinfo {author} {\bibfnamefont {D.}~\bibnamefont
  {Nettels}}, \bibinfo {author} {\bibfnamefont {B.~B.}\ \bibnamefont
  {Kragelund}}, \bibinfo {author} {\bibfnamefont {R.~B.}\ \bibnamefont
  {Best}},\ and\ \bibinfo {author} {\bibfnamefont {B.}~\bibnamefont
  {Schuler}},\ }\bibfield  {title} {\bibinfo {title} {Extreme disorder in an
  ultrahigh-affinity protein complex},\ }\href
  {https://doi.org/10.1038/nature25762} {\bibfield  {journal} {\bibinfo
  {journal} {Nature}\ }\textbf {\bibinfo {volume} {555}},\ \bibinfo {pages}
  {61–66} (\bibinfo {year} {2018})}\BibitemShut {NoStop}%
\bibitem [{\citenamefont {Jia}\ and\ \citenamefont
  {Muthukumar}(2022)}]{Jia2022}%
  \BibitemOpen
  \bibfield  {author} {\bibinfo {author} {\bibfnamefont {D.}~\bibnamefont
  {Jia}}\ and\ \bibinfo {author} {\bibfnamefont {M.}~\bibnamefont
  {Muthukumar}},\ }\bibfield  {title} {\bibinfo {title} {Dipole-driven
  interlude of mesomorphism in polyelectrolyte solutions},\ }\bibfield
  {journal} {\bibinfo  {journal} {Proceedings of the National Academy of
  Sciences}\ }\textbf {\bibinfo {volume} {119}},\ \href
  {https://doi.org/10.1073/pnas.2204163119} {10.1073/pnas.2204163119} (\bibinfo
  {year} {2022})\BibitemShut {NoStop}%
\bibitem [{\citenamefont {Liu}\ \emph {et~al.}(2023)\citenamefont {Liu},
  \citenamefont {Keum}, \citenamefont {Li}, \citenamefont {Chen}, \citenamefont
  {Hong}, \citenamefont {Wang}, \citenamefont {Sumpter}, \citenamefont
  {Advincula},\ and\ \citenamefont {Kumar}}]{Liu2023}%
  \BibitemOpen
  \bibfield  {author} {\bibinfo {author} {\bibfnamefont {Z.}~\bibnamefont
  {Liu}}, \bibinfo {author} {\bibfnamefont {J.~K.}\ \bibnamefont {Keum}},
  \bibinfo {author} {\bibfnamefont {T.}~\bibnamefont {Li}}, \bibinfo {author}
  {\bibfnamefont {J.}~\bibnamefont {Chen}}, \bibinfo {author} {\bibfnamefont
  {K.}~\bibnamefont {Hong}}, \bibinfo {author} {\bibfnamefont {Y.}~\bibnamefont
  {Wang}}, \bibinfo {author} {\bibfnamefont {B.~G.}\ \bibnamefont {Sumpter}},
  \bibinfo {author} {\bibfnamefont {R.}~\bibnamefont {Advincula}},\ and\
  \bibinfo {author} {\bibfnamefont {R.}~\bibnamefont {Kumar}},\ }\bibfield
  {title} {\bibinfo {title} {Anti-polyelectrolyte and polyelectrolyte effects
  on conformations of polyzwitterionic chains in dilute aqueous solutions},\
  }\bibfield  {journal} {\bibinfo  {journal} {PNAS Nexus}\ }\textbf {\bibinfo
  {volume} {2}},\ \href {https://doi.org/10.1093/pnasnexus/pgad204}
  {10.1093/pnasnexus/pgad204} (\bibinfo {year} {2023})\BibitemShut {NoStop}%
\bibitem [{\citenamefont {Di~Cera}\ and\ \citenamefont
  {Chen}(1993)}]{Di_Cera1993-kv}%
  \BibitemOpen
  \bibfield  {author} {\bibinfo {author} {\bibfnamefont {E.}~\bibnamefont
  {Di~Cera}}\ and\ \bibinfo {author} {\bibfnamefont {Z.}~\bibnamefont {Chen}},\
  }\bibfield  {title} {\bibinfo {title} {The binding capacity is a probability
  density function},\ }\href {https://doi.org/10.1016/s0006-3495(93)81033-6}
  {\bibfield  {journal} {\bibinfo  {journal} {Biophysical Journal}\ }\textbf
  {\bibinfo {volume} {65}},\ \bibinfo {pages} {164–170} (\bibinfo {year}
  {1993})}\BibitemShut {NoStop}%
\bibitem [{\citenamefont {Zhou}(2002)}]{Zhou2002}%
  \BibitemOpen
  \bibfield  {author} {\bibinfo {author} {\bibfnamefont {H.-X.}\ \bibnamefont
  {Zhou}},\ }\bibfield  {title} {\bibinfo {title} {A gaussian-chain model for
  treating residual charge–charge interactions in the unfolded state of
  proteins},\ }\href {https://doi.org/10.1073/pnas.052030599} {\bibfield
  {journal} {\bibinfo  {journal} {Proceedings of the National Academy of
  Sciences}\ }\textbf {\bibinfo {volume} {99}},\ \bibinfo {pages} {3569–3574}
  (\bibinfo {year} {2002})}\BibitemShut {NoStop}%
\bibitem [{\citenamefont {Zhou}(2003)}]{Zhou2003}%
  \BibitemOpen
  \bibfield  {author} {\bibinfo {author} {\bibfnamefont {H.-X.}\ \bibnamefont
  {Zhou}},\ }\bibfield  {title} {\bibinfo {title} {Direct test of the
  gaussian-chain model for treating residual charge-charge interactions in the
  unfolded state of proteins},\ }\href {https://doi.org/10.1021/ja0298491}
  {\bibfield  {journal} {\bibinfo  {journal} {Journal of the American Chemical
  Society}\ }\textbf {\bibinfo {volume} {125}},\ \bibinfo {pages} {2060–2061}
  (\bibinfo {year} {2003})}\BibitemShut {NoStop}%
\bibitem [{\citenamefont {Bhattacharjee}\ and\ \citenamefont
  {Muthukumar}(1987)}]{Bhattacharjee1987}%
  \BibitemOpen
  \bibfield  {author} {\bibinfo {author} {\bibfnamefont {S.~M.}\ \bibnamefont
  {Bhattacharjee}}\ and\ \bibinfo {author} {\bibfnamefont {M.}~\bibnamefont
  {Muthukumar}},\ }\bibfield  {title} {\bibinfo {title} {Statistical mechanics
  of solutions of semiflexible chains: A path integral formulation},\ }\href
  {https://doi.org/10.1063/1.452579} {\bibfield  {journal} {\bibinfo  {journal}
  {The Journal of Chemical Physics}\ }\textbf {\bibinfo {volume} {86}},\
  \bibinfo {pages} {411–418} (\bibinfo {year} {1987})}\BibitemShut {NoStop}%
\bibitem [{\citenamefont {Ghosh}\ \emph {et~al.}(2001)\citenamefont {Ghosh},
  \citenamefont {Carri},\ and\ \citenamefont {Muthukumar}}]{Ghosh2001}%
  \BibitemOpen
  \bibfield  {author} {\bibinfo {author} {\bibfnamefont {K.}~\bibnamefont
  {Ghosh}}, \bibinfo {author} {\bibfnamefont {G.~A.}\ \bibnamefont {Carri}},\
  and\ \bibinfo {author} {\bibfnamefont {M.}~\bibnamefont {Muthukumar}},\
  }\bibfield  {title} {\bibinfo {title} {Configurational properties of a single
  semiflexible polyelectrolyte},\ }\href {https://doi.org/10.1063/1.1386924}
  {\bibfield  {journal} {\bibinfo  {journal} {The Journal of Chemical Physics}\
  }\textbf {\bibinfo {volume} {115}},\ \bibinfo {pages} {4367–4375} (\bibinfo
  {year} {2001})}\BibitemShut {NoStop}%
\bibitem [{\citenamefont {Das}\ and\ \citenamefont {Pappu}(2013)}]{Das2013}%
  \BibitemOpen
  \bibfield  {author} {\bibinfo {author} {\bibfnamefont {R.~K.}\ \bibnamefont
  {Das}}\ and\ \bibinfo {author} {\bibfnamefont {R.~V.}\ \bibnamefont
  {Pappu}},\ }\bibfield  {title} {\bibinfo {title} {Conformations of
  intrinsically disordered proteins are influenced by linear sequence
  distributions of oppositely charged residues},\ }\href
  {https://doi.org/10.1073/pnas.1304749110} {\bibfield  {journal} {\bibinfo
  {journal} {Proceedings of the National Academy of Sciences}\ }\textbf
  {\bibinfo {volume} {110}},\ \bibinfo {pages} {13392–13397} (\bibinfo {year}
  {2013})}\BibitemShut {NoStop}%
\bibitem [{\citenamefont {Sing}\ and\ \citenamefont {Perry}(2020)}]{sing2020}%
  \BibitemOpen
  \bibfield  {author} {\bibinfo {author} {\bibfnamefont {C.~E.}\ \bibnamefont
  {Sing}}\ and\ \bibinfo {author} {\bibfnamefont {S.~L.}\ \bibnamefont
  {Perry}},\ }\bibfield  {title} {\bibinfo {title} {Recent progress in the
  science of complex coacervation},\ }\href
  {https://doi.org/10.1039/D0SM00001A} {\bibfield  {journal} {\bibinfo
  {journal} {Soft Matter}\ }\textbf {\bibinfo {volume} {16}},\ \bibinfo {pages}
  {2885} (\bibinfo {year} {2020})}\BibitemShut {NoStop}%
\bibitem [{\citenamefont {Choi}\ and\ \citenamefont
  {Pappu}(2018)}]{Choi2019-BASICS}%
  \BibitemOpen
  \bibfield  {author} {\bibinfo {author} {\bibfnamefont {J.-M.}\ \bibnamefont
  {Choi}}\ and\ \bibinfo {author} {\bibfnamefont {R.~V.}\ \bibnamefont
  {Pappu}},\ }\bibfield  {title} {\bibinfo {title} {Experimentally derived and
  computationally optimized backbone conformational statistics for blocked
  amino acids},\ }\href {https://doi.org/10.1021/acs.jctc.8b00572} {\bibfield
  {journal} {\bibinfo  {journal} {Journal of Chemical Theory and Computation}\
  }\textbf {\bibinfo {volume} {15}},\ \bibinfo {pages} {1355–1366} (\bibinfo
  {year} {2018})}\BibitemShut {NoStop}%
\bibitem [{\citenamefont {Zeng}\ \emph {et~al.}(2022)\citenamefont {Zeng},
  \citenamefont {Ruff},\ and\ \citenamefont {Pappu}}]{Zeng2022-jz}%
  \BibitemOpen
  \bibfield  {author} {\bibinfo {author} {\bibfnamefont {X.}~\bibnamefont
  {Zeng}}, \bibinfo {author} {\bibfnamefont {K.~M.}\ \bibnamefont {Ruff}},\
  and\ \bibinfo {author} {\bibfnamefont {R.~V.}\ \bibnamefont {Pappu}},\
  }\bibfield  {title} {\bibinfo {title} {Competing interactions give rise to
  two-state behavior and switch-like transitions in charge-rich intrinsically
  disordered proteins},\ }\bibfield  {journal} {\bibinfo  {journal}
  {Proceedings of the National Academy of Sciences}\ }\textbf {\bibinfo
  {volume} {119}},\ \href {https://doi.org/10.1073/pnas.2200559119}
  {10.1073/pnas.2200559119} (\bibinfo {year} {2022})\BibitemShut {NoStop}%
\bibitem [{\citenamefont {Cheng}\ \emph {et~al.}(2010)\citenamefont {Cheng},
  \citenamefont {Cetinkaya},\ and\ \citenamefont {Gr\"{a}ter}}]{Cheng2010}%
  \BibitemOpen
  \bibfield  {author} {\bibinfo {author} {\bibfnamefont {S.}~\bibnamefont
  {Cheng}}, \bibinfo {author} {\bibfnamefont {M.}~\bibnamefont {Cetinkaya}},\
  and\ \bibinfo {author} {\bibfnamefont {F.}~\bibnamefont {Gr\"{a}ter}},\
  }\bibfield  {title} {\bibinfo {title} {How sequence determines elasticity of
  disordered proteins},\ }\href {https://doi.org/10.1016/j.bpj.2010.10.011}
  {\bibfield  {journal} {\bibinfo  {journal} {Biophysical Journal}\ }\textbf
  {\bibinfo {volume} {99}},\ \bibinfo {pages} {3863–3869} (\bibinfo {year}
  {2010})}\BibitemShut {NoStop}%
\bibitem [{\citenamefont {Martin}\ \emph {et~al.}(2016)\citenamefont {Martin},
  \citenamefont {Holehouse}, \citenamefont {Grace}, \citenamefont {Hughes},
  \citenamefont {Pappu},\ and\ \citenamefont {Mittag}}]{Martin2016}%
  \BibitemOpen
  \bibfield  {author} {\bibinfo {author} {\bibfnamefont {E.~W.}\ \bibnamefont
  {Martin}}, \bibinfo {author} {\bibfnamefont {A.~S.}\ \bibnamefont
  {Holehouse}}, \bibinfo {author} {\bibfnamefont {C.~R.}\ \bibnamefont
  {Grace}}, \bibinfo {author} {\bibfnamefont {A.}~\bibnamefont {Hughes}},
  \bibinfo {author} {\bibfnamefont {R.~V.}\ \bibnamefont {Pappu}},\ and\
  \bibinfo {author} {\bibfnamefont {T.}~\bibnamefont {Mittag}},\ }\bibfield
  {title} {\bibinfo {title} {Sequence determinants of the conformational
  properties of an intrinsically disordered protein prior to and upon multisite
  phosphorylation},\ }\href {https://doi.org/10.1021/jacs.6b10272} {\bibfield
  {journal} {\bibinfo  {journal} {Journal of the American Chemical Society}\
  }\textbf {\bibinfo {volume} {138}},\ \bibinfo {pages} {15323–15335}
  (\bibinfo {year} {2016})}\BibitemShut {NoStop}%
\bibitem [{\citenamefont {Holla}\ \emph {et~al.}(2024)\citenamefont {Holla},
  \citenamefont {Martin}, \citenamefont {Dannenhoffer-Lafage}, \citenamefont
  {Ruff}, \citenamefont {K\"{o}nig}, \citenamefont {N\"{u}esch}, \citenamefont
  {Chowdhury}, \citenamefont {Louis}, \citenamefont {Soranno}, \citenamefont
  {Nettels}, \citenamefont {Pappu}, \citenamefont {Best}, \citenamefont
  {Mittag},\ and\ \citenamefont {Schuler}}]{Holla2024}%
  \BibitemOpen
  \bibfield  {author} {\bibinfo {author} {\bibfnamefont {A.}~\bibnamefont
  {Holla}}, \bibinfo {author} {\bibfnamefont {E.~W.}\ \bibnamefont {Martin}},
  \bibinfo {author} {\bibfnamefont {T.}~\bibnamefont {Dannenhoffer-Lafage}},
  \bibinfo {author} {\bibfnamefont {K.~M.}\ \bibnamefont {Ruff}}, \bibinfo
  {author} {\bibfnamefont {S.~L.~B.}\ \bibnamefont {K\"{o}nig}}, \bibinfo
  {author} {\bibfnamefont {M.~F.}\ \bibnamefont {N\"{u}esch}}, \bibinfo
  {author} {\bibfnamefont {A.}~\bibnamefont {Chowdhury}}, \bibinfo {author}
  {\bibfnamefont {J.~M.}\ \bibnamefont {Louis}}, \bibinfo {author}
  {\bibfnamefont {A.}~\bibnamefont {Soranno}}, \bibinfo {author} {\bibfnamefont
  {D.}~\bibnamefont {Nettels}}, \bibinfo {author} {\bibfnamefont {R.~V.}\
  \bibnamefont {Pappu}}, \bibinfo {author} {\bibfnamefont {R.~B.}\ \bibnamefont
  {Best}}, \bibinfo {author} {\bibfnamefont {T.}~\bibnamefont {Mittag}},\ and\
  \bibinfo {author} {\bibfnamefont {B.}~\bibnamefont {Schuler}},\ }\bibfield
  {title} {\bibinfo {title} {Identifying sequence effects on chain dimensions
  of disordered proteins by integrating experiments and simulations},\ }\href
  {https://doi.org/10.1021/jacsau.4c00673} {\bibfield  {journal} {\bibinfo
  {journal} {JACS Au}\ }\textbf {\bibinfo {volume} {4}},\ \bibinfo {pages}
  {4729–4743} (\bibinfo {year} {2024})}\BibitemShut {NoStop}%
\bibitem [{\citenamefont {Vitalis}\ and\ \citenamefont
  {Pappu}(2008{\natexlab{b}})}]{Vitalis2009}%
  \BibitemOpen
  \bibfield  {author} {\bibinfo {author} {\bibfnamefont {A.}~\bibnamefont
  {Vitalis}}\ and\ \bibinfo {author} {\bibfnamefont {R.~V.}\ \bibnamefont
  {Pappu}},\ }\bibfield  {title} {\bibinfo {title} {Absinth: A new continuum
  solvation model for simulations of polypeptides in aqueous solutions},\
  }\href {https://doi.org/10.1002/jcc.21005} {\bibfield  {journal} {\bibinfo
  {journal} {Journal of Computational Chemistry}\ }\textbf {\bibinfo {volume}
  {30}},\ \bibinfo {pages} {673–699} (\bibinfo {year}
  {2008}{\natexlab{b}})}\BibitemShut {NoStop}%
\bibitem [{\citenamefont {Chandler}(1987)}]{Chandler1987}%
  \BibitemOpen
  \bibfield  {author} {\bibinfo {author} {\bibfnamefont {D.}~\bibnamefont
  {Chandler}},\ }\href@noop {} {\emph {\bibinfo {title} {Introduction to Modern
  Statistical Mechanics}}}\ (\bibinfo  {publisher} {Oxford University Press},\
  \bibinfo {year} {1987})\ p.\ \bibinfo {pages} {288}\BibitemShut {NoStop}%
\bibitem [{\citenamefont {Flory}\ and\ \citenamefont
  {Krigbaum}(1950)}]{flory1950}%
  \BibitemOpen
  \bibfield  {author} {\bibinfo {author} {\bibfnamefont {P.~J.}\ \bibnamefont
  {Flory}}\ and\ \bibinfo {author} {\bibfnamefont {W.~R.}\ \bibnamefont
  {Krigbaum}},\ }\bibfield  {title} {\bibinfo {title} {Statistical mechanics of
  dilute polymer solutions. {II}},\ }\href {https://doi.org/10.1063/1.1747866}
  {\bibfield  {journal} {\bibinfo  {journal} {JCP}\ }\textbf {\bibinfo {volume}
  {18}},\ \bibinfo {pages} {1086} (\bibinfo {year} {1950})}\BibitemShut
  {NoStop}%
\bibitem [{\citenamefont {Podgornik}(1993)}]{podgornik1993}%
  \BibitemOpen
  \bibfield  {author} {\bibinfo {author} {\bibfnamefont {R.}~\bibnamefont
  {Podgornik}},\ }\bibfield  {title} {\bibinfo {title} {A variational approach
  to charged polymer chains: Polymer mediated interactions},\ }\href
  {https://doi.org/10.1063/1.465439} {\bibfield  {journal} {\bibinfo  {journal}
  {JCP}\ }\textbf {\bibinfo {volume} {99}},\ \bibinfo {pages} {7221} (\bibinfo
  {year} {1993})}\BibitemShut {NoStop}%
\bibitem [{\citenamefont {Born}(1920)}]{Born1920-sk}%
  \BibitemOpen
  \bibfield  {author} {\bibinfo {author} {\bibfnamefont {M.}~\bibnamefont
  {Born}},\ }\bibfield  {title} {\bibinfo {title} {Volumen und
  hydratationsw\"{a}rme der ionen},\ }\href
  {https://doi.org/10.1007/bf01881023} {\bibfield  {journal} {\bibinfo
  {journal} {Zeitschrift f\"{u}r Physik}\ }\textbf {\bibinfo {volume} {1}},\
  \bibinfo {pages} {45–48} (\bibinfo {year} {1920})}\BibitemShut {NoStop}%
\bibitem [{\citenamefont {Schutz}\ and\ \citenamefont
  {Warshel}(2001)}]{Schutz2001}%
  \BibitemOpen
  \bibfield  {author} {\bibinfo {author} {\bibfnamefont {C.~N.}\ \bibnamefont
  {Schutz}}\ and\ \bibinfo {author} {\bibfnamefont {A.}~\bibnamefont
  {Warshel}},\ }\bibfield  {title} {\bibinfo {title} {What are the dielectric
  “constants” of proteins and how to validate electrostatic models?},\
  }\href {https://doi.org/10.1002/prot.1106} {\bibfield  {journal} {\bibinfo
  {journal} {Proteins: Structure, Function, and Bioinformatics}\ }\textbf
  {\bibinfo {volume} {44}},\ \bibinfo {pages} {400–417} (\bibinfo {year}
  {2001})}\BibitemShut {NoStop}%
\bibitem [{\citenamefont {Chowdhury}\ \emph {et~al.}(2019)\citenamefont
  {Chowdhury}, \citenamefont {Kovalenko}, \citenamefont {Aramburu},
  \citenamefont {Tan}, \citenamefont {Ernsting},\ and\ \citenamefont
  {Lemke}}]{Chowdhury2019}%
  \BibitemOpen
  \bibfield  {author} {\bibinfo {author} {\bibfnamefont {A.}~\bibnamefont
  {Chowdhury}}, \bibinfo {author} {\bibfnamefont {S.~A.}\ \bibnamefont
  {Kovalenko}}, \bibinfo {author} {\bibfnamefont {I.~V.}\ \bibnamefont
  {Aramburu}}, \bibinfo {author} {\bibfnamefont {P.~S.}\ \bibnamefont {Tan}},
  \bibinfo {author} {\bibfnamefont {N.~P.}\ \bibnamefont {Ernsting}},\ and\
  \bibinfo {author} {\bibfnamefont {E.~A.}\ \bibnamefont {Lemke}},\ }\bibfield
  {title} {\bibinfo {title} {Mechanism‐dependent modulation of ultrafast
  interfacial water dynamics in intrinsically disordered protein complexes},\
  }\href {https://doi.org/10.1002/anie.201813354} {\bibfield  {journal}
  {\bibinfo  {journal} {Angewandte Chemie International Edition}\ }\textbf
  {\bibinfo {volume} {58}},\ \bibinfo {pages} {4720–4724} (\bibinfo {year}
  {2019})}\BibitemShut {NoStop}%
\bibitem [{\citenamefont {Makhatadze}(2017)}]{Makhatadze2017}%
  \BibitemOpen
  \bibfield  {author} {\bibinfo {author} {\bibfnamefont {G.~I.}\ \bibnamefont
  {Makhatadze}},\ }\bibfield  {title} {\bibinfo {title} {Linking computation
  and experiments to study the role of charge–charge interactions in protein
  folding and stability},\ }\href
  {https://doi.org/10.1088/1478-3975/14/1/013002} {\bibfield  {journal}
  {\bibinfo  {journal} {Physical Biology}\ }\textbf {\bibinfo {volume} {14}},\
  \bibinfo {pages} {013002} (\bibinfo {year} {2017})}\BibitemShut {NoStop}%
\bibitem [{\citenamefont {Chen}\ and\ \citenamefont {Pappu}(2007)}]{Chen2007}%
  \BibitemOpen
  \bibfield  {author} {\bibinfo {author} {\bibfnamefont {A.~A.}\ \bibnamefont
  {Chen}}\ and\ \bibinfo {author} {\bibfnamefont {R.~V.}\ \bibnamefont
  {Pappu}},\ }\bibfield  {title} {\bibinfo {title} {Quantitative
  characterization of ion pairing and cluster formation in strong 1:1
  electrolytes},\ }\href {https://doi.org/10.1021/jp0708547} {\bibfield
  {journal} {\bibinfo  {journal} {The Journal of Physical Chemistry B}\
  }\textbf {\bibinfo {volume} {111}},\ \bibinfo {pages} {6469–6478} (\bibinfo
  {year} {2007})}\BibitemShut {NoStop}%
\bibitem [{\citenamefont {Muthukumar}(2002)}]{muthu2002}%
  \BibitemOpen
  \bibfield  {author} {\bibinfo {author} {\bibfnamefont {M.}~\bibnamefont
  {Muthukumar}},\ }\bibfield  {title} {\bibinfo {title} {Phase diagram of
  polyelectrolyte solutions:{\hspace{0.167em}} weak polymer effect},\ }\href
  {https://doi.org/10.1021/ma021456z} {\bibfield  {journal} {\bibinfo
  {journal} {Macromolecules}\ }\textbf {\bibinfo {volume} {35}},\ \bibinfo
  {pages} {9142} (\bibinfo {year} {2002})}\BibitemShut {NoStop}%
\bibitem [{\citenamefont {McQuarrie}(2000)}]{mcquarrie2000}%
  \BibitemOpen
  \bibfield  {author} {\bibinfo {author} {\bibfnamefont {D.}~\bibnamefont
  {McQuarrie}},\ }\href {https://books.google.co.in/books?id=itcpPnDnJM0C}
  {\emph {\bibinfo {title} {Statistical Mechanics}}}\ (\bibinfo  {publisher}
  {University Science Books},\ \bibinfo {year} {2000})\BibitemShut {NoStop}%
\bibitem [{\citenamefont {Kudlay}\ \emph {et~al.}(2004)\citenamefont {Kudlay},
  \citenamefont {Ermoshkin},\ and\ \citenamefont {Olvera de~la
  Cruz}}]{delacruz2004}%
  \BibitemOpen
  \bibfield  {author} {\bibinfo {author} {\bibfnamefont {A.}~\bibnamefont
  {Kudlay}}, \bibinfo {author} {\bibfnamefont {A.~V.}\ \bibnamefont
  {Ermoshkin}},\ and\ \bibinfo {author} {\bibfnamefont {M.}~\bibnamefont
  {Olvera de~la Cruz}},\ }\bibfield  {title} {\bibinfo {title} {Complexation of
  oppositely charged polyelectrolytes: Effect of ion pair formation},\ }\href
  {https://doi.org/10.1021/ma048519t} {\bibfield  {journal} {\bibinfo
  {journal} {Macromolecules}\ }\textbf {\bibinfo {volume} {37}},\ \bibinfo
  {pages} {9231} (\bibinfo {year} {2004})},\ \Eprint
  {https://arxiv.org/abs/https://doi.org/10.1021/ma048519t}
  {https://doi.org/10.1021/ma048519t} \BibitemShut {NoStop}%
\bibitem [{\citenamefont {Nelder}\ and\ \citenamefont
  {Mead}(1965)}]{Nelder1965}%
  \BibitemOpen
  \bibfield  {author} {\bibinfo {author} {\bibfnamefont {J.~A.}\ \bibnamefont
  {Nelder}}\ and\ \bibinfo {author} {\bibfnamefont {R.}~\bibnamefont {Mead}},\
  }\bibfield  {title} {\bibinfo {title} {A simplex method for function
  minimization},\ }\href {https://doi.org/10.1093/comjnl/7.4.308} {\bibfield
  {journal} {\bibinfo  {journal} {The Computer Journal}\ }\textbf {\bibinfo
  {volume} {7}},\ \bibinfo {pages} {308–313} (\bibinfo {year}
  {1965})}\BibitemShut {NoStop}%
\bibitem [{\citenamefont {Bawendi}\ and\ \citenamefont
  {Freed}(1985)}]{Bawendi1985}%
  \BibitemOpen
  \bibfield  {author} {\bibinfo {author} {\bibfnamefont {M.~G.}\ \bibnamefont
  {Bawendi}}\ and\ \bibinfo {author} {\bibfnamefont {K.~F.}\ \bibnamefont
  {Freed}},\ }\bibfield  {title} {\bibinfo {title} {A wiener integral model for
  stiff polymer chains},\ }\href {https://doi.org/10.1063/1.449296} {\bibfield
  {journal} {\bibinfo  {journal} {The Journal of Chemical Physics}\ }\textbf
  {\bibinfo {volume} {83}},\ \bibinfo {pages} {2491–2496} (\bibinfo {year}
  {1985})}\BibitemShut {NoStop}%
\bibitem [{\citenamefont {Lagowski}\ \emph {et~al.}(1991)\citenamefont
  {Lagowski}, \citenamefont {Noolandi},\ and\ \citenamefont
  {Nickel}}]{Lagowski1991}%
  \BibitemOpen
  \bibfield  {author} {\bibinfo {author} {\bibfnamefont {J.~B.}\ \bibnamefont
  {Lagowski}}, \bibinfo {author} {\bibfnamefont {J.}~\bibnamefont {Noolandi}},\
  and\ \bibinfo {author} {\bibfnamefont {B.}~\bibnamefont {Nickel}},\
  }\bibfield  {title} {\bibinfo {title} {Stiff chain model—functional
  integral approach},\ }\href {https://doi.org/10.1063/1.461106} {\bibfield
  {journal} {\bibinfo  {journal} {The Journal of Chemical Physics}\ }\textbf
  {\bibinfo {volume} {95}},\ \bibinfo {pages} {1266–1269} (\bibinfo {year}
  {1991})}\BibitemShut {NoStop}%
\end{thebibliography}%
\clearpage
\appendix

\section{Evaluation of $F_{pol}$}\label{appendix:evaluation_of_F5}

	A variational procedure \cite{muthu1987}, that involves a trial Hamiltonian based on the unperturbed Gaussian chain, is used to recast the Hamiltonian of Eq. \ref{eq:H} as 
	\begin{align}
		\mathcal{H}=&\mathcal{H}_{\text{trial}}+(\mathcal{H}-\mathcal{H}_{\text{trial}}),
	\end{align}
	where
	\begin{align}\label{trial-Hamiltonian}
		\mathcal{H}_{\text{trial}}=\frac{3}{2 \ell\ell_{1}}\int_{0}^{N} d s\left(\frac{\partial \mathbf{R}\left(s\right)}{\partial s}\right)^{2}.
	\end{align}
Here, $\ell_{1}$ represents the variational parameter that characterizes the effective expansion factor of the polyion in comparison to its unperturbed Gaussian chain size\cite{edwards1979, muthu1987, muthu2004, Muthukumar2023-book}.
	The mean-field assumption is based on the Gibbs-Bogoliubov-Feynman inequality \cite{Chandler1987},
	\begin{align}
		\left\langle\mathrm{e}^{-\beta \mathcal{H}}\right\rangle_{\mathcal{H}_\mathrm{{trial}}} \geq \mathrm{e}^{-\beta\left\langle \mathcal{H}\right\rangle_{\mathcal{H}_\mathrm{{trial}}}};\label{avgexample}
	\end{align}
	This implies that the free energy,
	\begin{align} \label{F5-trial}
		\widetilde{F}_5=&\langle \beta(\mathcal{H}_0-\mathcal{H}_\mathrm{{trial}})\rangle_{\mathcal{H}_\mathrm{{trial}}}+\langle \beta \mathcal{H}_{ex}\rangle_{\mathcal{H}_\mathrm{{trial}}}\nonumber\\
		&+\langle \beta \mathcal{H}_{el}\rangle_{\mathcal{H}_\mathrm{{trial}}},
	\end{align}
	needs to be extremized with respect to the charge ($ f $) and size (expansion factor, $ \ell_{1} $) of the  poly-acid. Note that $f =  N_c/N  - M_{\mathrm p}/N - M_{s}/N$ and it is influenced by the total number of ionizable moieties, taken to be the chain length $N$, the number of acidic sites where protons are bound $M_{\mathrm p}$, and the number of condensed salt ions $M_{s}$. If one shifts to the polymer coordinate (the spatial coordinate ${\bf r}$) the Hamiltonian can be approximately recast in terms of the monomer density profile of the poly-acid \cite{flory1950,podgornik1993,Muthukumar2012,Ghosh2023}. If one assumes a Gaussian distribution, the chain density centered at $\mathbf{r}_{0}$ and positioned at $\mathbf{r}$ can be expressed as
	\begin{align}\label{rho-def}
		&\rho(\mathbf{r})=N\left(\frac{3}{4 \pi R_{g}^{2}}\right)^{3 / 2} \exp \left[-\frac{3 (\left|\mathbf{r}-\mathbf{r}_{0}\right|)^{2}}{2 R_{g}^{2}}\right].
	\end{align}
	Under the assumption of uniform expansion of the poly-acid \cite{edwards1979,muthu1987,muthu2004,Muthukumar2023-book}, the average dimensionless radius of gyration of the chain can be obtained as \cref{rg_def}.
    \begin{align}\label{rg_def}
		\widetilde{R}_{g}=\sqrt{\frac{N \widetilde{\ell}_{1}}{6}},
	\end{align}
	where $\widetilde{R}_{g}=R_g/\ell$ and $\widetilde{\ell}_{1}=\ell_{1}/\ell$.
	
	Using the Fourier transform (in ${\bf k}$-space) of the monomer density profile (Eq. \ref{rho-def}) and integrating the averaged interaction of monomers (Eq. \ref{F5-trial}), the total free energy contribution due to the polymer degrees of freedom included in the Hamiltonian (Eq. \ref{eq:H}) can be obtained in the form given by \cref{F5-appendix}.

\begin{widetext}
	\section{Mean structure factor in the ensemble of the unperturbed chain}\label{appendix:structure-factor}
	
		$\mathbf{R}(s)$ can be written as the inverse Fourier transform of $\mathbf{R}(q)$ as
		\begin{align}\label{rs2rq}
				\mathbf{R}(s)=\int_{-\infty}^{\infty} \frac{d q}{2 \pi} \mathbf{R}(q) \exp (i q s).
		\end{align}
	We take the trial Hamiltonian of a chain and representing it using the Fourier transform (given in Eq.\eqref{rs2rq}), we obtain
	\begin{align}
		\beta \mathcal{H}_{0}(\{\mathbf{R}\left(q\right)\})=& \int_{-\infty}^{\infty} \frac{d q}{2 \pi} \frac{R_{q_{1}}^2}{g(q)}
	\end{align}
	with $ g(q)={2 \ell_{1}}/{(3 q^{2})} $. Here, we follow the exact steps as outlined in ref. \cite{muthu1987}. Using the definitions in Eqs. \eqref{avgexample} and \eqref{rs2rq}, we can discretize the notation and obtain
		\begin{align}\label{Exp-avg}
			&\left\langle\exp\left\{i \mathbf{k} \cdot \int_{-\infty}^{\infty} \frac{d q^{\prime}}{2 \pi} \mathbf{R}\left(q^{\prime}\right)\left[\exp \left(i q^{\prime} s\right)-\exp \left(i q^{\prime} s^{\prime}\right)\right]\right\}\right\rangle_0 \\
			=& \frac{\int \prod_p d \mathbf{R}(p) \exp \left\{-\sum_{p=-\infty}^{\infty} \frac{\mathbf{R}^2(p)}{L g}+\frac{i \mathbf{k}}{L} \cdot \sum_{p=-\infty}^{\infty} \mathbf{R}(p)\left[\exp \left(\frac{2 \pi i p s}{L}\right)-\exp \left(\frac{2 \pi i p s^{\prime}}{L}\right)\right]\right\}}{\int \prod_p d \mathbf{R}(p) \exp \left[-\sum_{p=-\infty}^{\infty} \frac{\mathbf{R}^2(p)}{L g}\right]} \\
			=& \prod_p \frac{\int d \mathbf{R}(p) \exp \left\{-\frac{\mathbf{R}^2(p)}{L g}+\frac{i \mathbf{k}}{L} \cdot \mathbf{R}(p)\left[\exp \left(\frac{2 \pi i p s}{L}\right)-\exp \left(\frac{2 \pi i p s^{\prime}}{L}\right)\right]\right\}}{2 \pi(p) \exp \left[-\mathbf{R}^2(p) / L g\right]} \\
			=& \exp \left[-k^2 \int_{-\infty}^{\infty} \frac{d q}{2 \pi} g(q) \sin ^2\left(\frac{q\left|s-s^{\prime}\right|}{2}\right) \right].
		\end{align}
	\section{The single chain density profile}\label{appendix-section:rho-fourier}
	Following \cite{Ghosh2025}, the density profile of a single chain, $\rho(\mathbf{r})$, is defined as $\int d s \delta\left(\mathbf{R}(s)-\mathbf{r}\right)$ in Eq.\eqref{rho-ito-Rsi}. We assume that $\rho(\mathbf{r})$ can be approximated by a Gaussian distribution \cite{flory1950, Ghosh2025}, and at a position $\mathbf{r}$, the density profile is given by:
	\begin{align}\label{rho-def-determination-of-rho}
		\rho(\mathbf{r}) &=A e^{-a \mathbf{(\left|\ro-\ro_{0}\right|)}^{2}},
	\end{align}
	where $A$ and $a$ are determined based on the definitions $N$ and $ R_{g} $:
	\begin{align}\label{N-def-determination-of-rho}
		&{N}=\int_{0}^{\infty} 4 \pi \mathbf{r}^{2} \rho(\mathbf{r})d\mathbf{r}=\frac{\pi^{3 / 2} A}{a^{3 / 2}},
	\end{align}
	\begin{align}\label{Rg-def-determination-of-rho}
		&R_{g}^{2}=\frac{\int_{0}^{\infty} 4 \pi \mathbf{r}^{4} \rho(\mathbf{r}) d \mathbf{r}}{\int 4 \pi \mathbf{r}^{2} \rho(\mathbf{r}) d \mathbf{r}}=\frac{3 \pi^{3 / 2} A}{2 N a^{5 / 2}},
	\end{align}
	respectively. Solving Eq.\eqref{N-def-determination-of-rho} and\eqref{Rg-def-determination-of-rho}, we get
	\begin{align}
		&a=\frac{3}{2 R_{g}^{2}},\qquad
		A=N\left(\frac{3}{2 \pi R_{g}^{2}}\right)^{3 / 2}.
	\end{align}
	Therefore, by substituting these values of $a$ and $A$ back into Eq.\eqref{rho-def-determination-of-rho}, we obtain:
	\begin{align}
		&\rho(\ro)=N\left(\frac{3}{4 \pi R_{g}^{2}}\right)^{3 / 2} \exp \left[-\frac{3 (\left|\ro-\ro_{0}\right|)^{2}}{2 R_{g}^{2}}\right].
		\label{rho-appendix}
	\end{align}
	The Fourier transform of $ \rho(\ro) $ is given by 
	\begin{align}\label{conjugate-rho}
		\mathcal{F}[\rho(\mathbf{r})]=\hat{\rho}(\mathbf{k})N \exp{\left(-\frac{k^{2} R_{g}^{2}}{6}\right)} .
	\end{align}
		\section{Derivation of $ \mathcal{H}_{ex} $ and $\mathcal{H}_{el} $}\label{appendix-section:Hex}
		To calculate the averaged excluded volume interaction, we need to move to the polymer coordinate system. We start by considering the intra-chain interaction term (see the last term in Eq. \eqref{eq:H_ex}) as,
		%
		\begin{align}
			\mathcal{H}_{ex}\beta&=2\omega\ell^{3} \int_{0}^{N} ds \int_{0}^{N} ds^{\prime} \delta(\mathbf{R}(s)-\mathbf{R}(s^{\prime}))\nonumber\\
			&=w_{12}\ell^{3}\int d s \int d s^{\prime} \left[\int d^{3} \mathbf{r} \int d^{3} \mathbf{r}^{\prime} \delta\left(\mathbf{R}\left(s\right)-\mathbf{r}\right) \delta\left(\mathbf{r}-\mathbf{r}^{\prime}\right)\delta\left(\mathbf{R}\left(s^{\prime}\right)-\mathbf{r}^{\prime}\right)\right]\nonumber\\
			&=w_{12}\ell^{3}\int d^{3} \mathbf{r} \int d^{3} \mathbf{r}^{\prime} \left[\int d s_{1} \delta(\mathbf{R}(s)-\mathbf{r}) \delta(\mathbf{r}-\mathbf{r}^{\prime})\int d s^{\prime} \delta(\mathbf{R}(s^{\prime})-\mathbf{r}^\prime)\right].
		\end{align}
		In the above equation, we define, 
		\begin{align}\label{rho-ito-Rsi}
			\rho(\mathbf{r})=\int d s \delta\left(\mathbf{R}(s)-\mathbf{r}\right).
		\end{align}
		where $\rho(\mathbf{r})$ is the density of the polymer at position $\mathbf{r}$.
		For the of chain which avoids self intersection due to excluded volume effects, the above equation becomes, 
		\begin{align}\label{intra-chain-H-ex}
			&\mathcal{H}_{ex}\beta=\omega\ell^{3}\int \frac{d^{3} \mathbf{k}}{(2 \pi)^{3}} \left|\hat{\rho}(\mathbf{k})\right|^{2}
		\end{align}
		Using our knowledge of the single chain density profile (given by Eq. \eqref{rho-appendix}), we rewrite the electrostatic interaction in Fourier space. The Fourier transform of $ 	U_{ij}\left(\mathbf{r}-\mathbf{r}^{\prime}\right) $ can be obtained as,
		\begin{align}\label{conjugate-Uij}
			\hat{U}_{ij}^{el}(\mathbf{k})=\frac{f_{i}f_{j}\ell_{B}}{2}\frac{4\pi}{\mathbf{k}^2+\kappa^2}.
		\end{align}
		Using Eqs. \eqref{conjugate-Uij}, \eqref{rho-appendix} and assuming $ \widetilde{R}_{g} $ is encompasses contributions from the excluded volume effect for the chain, we write the intra-chain electrostatic interaction as,
		\begin{align}\label{inter-chain-H-el}
			\mathcal{H}_{el}(\mathbf{R})\beta=&\int d^{3} \mathbf{r}^{\prime} \int d^{3} \mathbf{r} \quad \rho\left(\mathbf{r}^{\prime}\right) U_{11}^{el}\left(\mathbf{r}-\mathbf{r}^{\prime}\right) \rho(\mathbf{r}) =\int d^{3} \mathbf{r}^{\prime} \int d^{3} \mathbf{r} \rho\left(\mathbf{r}^{\prime}\right)\left[\int \frac{d^{3} \mathbf{k}}{(2 \pi)^{3}} \hat{U}_{11}^{el}(\mathbf{k}) e^{-i \mathbf{k} \cdot\left(\mathbf{r}-\mathbf{r}^{\prime}\right)}\right] \rho(\mathbf{r})\nonumber\\
					&=\int \frac{d^{3} \mathbf{k}}{(2 \pi)^{3}}\left[\int d^{3} \mathbf{r}^{\prime} \rho\left(\mathbf{r}^{\prime}\right) e^{i \mathbf{k}. \mathbf{r}^{\prime}}\hat{U}_{11}^{el}(\mathbf{k}) \int d^{3} \mathbf{r} \rho(\mathbf{r}) e^{-i \mathbf{k}. \mathbf{r}}\right]\nonumber\\
			&=\int \frac{d^{3} \mathbf{k}}{(2 \pi)^{3}} \hat{\rho}(\mathbf{k}) \hat{U}_{11}^{el}(\mathbf{k}) \hat{\rho}(-\mathbf{k}) e^{-i \mathbf{k}\cdot \mathbf{R}}.
		\end{align}	
		Accordingly, the intra-chain electrostatic interaction can be obtained as 
		\begin{align}\label{intra-chain-H-el}
			\mathcal{H}_{el}(\mathbf{R})\beta=\int \frac{d^{3} \mathbf{k}}{(2 \pi)^{3}} \left|\hat{\rho}(\mathbf{k})\right|^{2}\hat{U}_{11}^{el}(\mathbf{k}).
		\end{align}
\end{widetext}

\section{Binding constant for the salt ions}\label{appendix:binding_constant}
 The free energy required to charge a protein side chain (such as Asp$^-$ or Glu$^-$) of radius $a_1$ in a medium with dielectric constant $\varepsilon$ is given by the Born energy for charging the ionic species in a dilectric continuum \cite{Born1920-sk},
 ${e^2}/{8\pi \varepsilon_0 \varepsilon a_1}.$
Similarly, for a counterion (e.g., Na$^+$ or Cl$^-$) of radius $a_2$, the Born energy is, ${e^2}/{8\pi \varepsilon_0 \varepsilon a_2}.$
For proteins, the local dielectric constant $\varepsilon_{\ell}$ near the charged side chain is thought to lower than that of bulk water \cite{Schutz2001}. The relevant values are highly context-dependent and have been estimated mainly for folded proteins as opposed to intrinsically disordered domains.  Recent experiments \cite{Chowdhury2019} and theoretical estimates \cite{Muthukumar2023-book} suggest that $\varepsilon_{\ell}$ may range from $\sim5$ in the vicinity of a protein surface to $\sim 80$ in the bulk solvent, over nanometer length scales. In the following, $\varepsilon_{\ell}$ is treated as an effective parameter to capture local dielectric inhomogeneities \cite{Makhatadze2017}.

The electrostatic binding free energy between a charged side chain and a counterion in a continuum at distance $r$ is
\[
-\,\frac{e^2}{4\pi \varepsilon_0 \varepsilon_{\ell} r}.
\]
Thus, the free energy change for forming an ion pair (e.g., Glu$^-$–Na$^+$) is
\begin{align}
    \frac{\Delta G}{k_B T}
    &= -\frac{e^2}{4\pi \varepsilon_0 \varepsilon_{\ell} r}
    - \frac{e^2}{8\pi \varepsilon_0}
    \left(
        \frac{1}{\varepsilon a_1}
        + \frac{1}{\varepsilon_{\ell} a_2}
    \right)\nonumber\\
    &\equiv -\frac{e^2}{4\pi\varepsilon_0 \varepsilon_{\ell} d},
\end{align}
where,
\[
\frac{1}{d} = \frac{1}{r} + \frac{1}{2a_2} + \frac{\varepsilon_{\ell}}{2\varepsilon a_1}.
\]
Both $\varepsilon_{\ell}$ and $d$ depend on the specific side chain chemistry, the local structure of the poly-acid, and the type of counterion involved. Following Ref.~\cite{muthu2017,Chowdhury2023,Phillips2024,Ghosh2025}, these contributions can be recast using a parameter $\delta/d = \varepsilon / (\varepsilon_{\ell} d)$, yielding
\begin{equation}
    \frac{\Delta G}{k_B T} = -\frac{\ell_B}{d\ell_{0}}\,\delta.
\end{equation}

\section{Free energy components}\label{appendix:free_energy_components}

\mycomment{\begin{align}
\mathrm{NaCl} &\rightleftharpoons \mathrm{Na}^{+} + \mathrm{Cl}^{-}, \nonumber\qquad
\mathrm{HCl}  &\rightleftharpoons \mathrm{H}^{+} + \mathrm{Cl}^{-}.\nonumber
\end{align}}

We consider a solution containing monovalent salt, which we assume to be fully dissociated. For the values of $c_s$ that we consider in this work, this assumption is reasonable \cite{Chen2007}. An overbar denotes number densities that are rendered dimensionless using the length scale $r_c$ and dividing the relevant parameters by the reduced volume $\Omega/r_c^3$ (definitions below). We also define the degree of ionization as, $ f=(N_c-M_p-M_s)/N $, where $N_c,~M_p,~M_s,$ and $N$ are the number of ionizable monomers, number of protonated sites, number of condensed ions, and number of total monomers respectively, $ k_{B} $ is the Boltzmann constant, $  T $ is the absolute temperature. The  chain density $\rho$ is rewritten as $\bar{\rho}=N/(\Omega / r_c^3)=(N \ell_{0}^{3}/\Omega)(r_c^{3}/\ell_{0}^{3})=\tilde{\rho} \tilde{r}_{c}^3$; $ \bar{c}_s $ is defined as $ \bar{c}_s=n_{s} /(\Omega / r_c^3)=(n_s \ell_{0}^{3}/\Omega)(r_c^{3}/\ell_{0}^{3})=\tilde{c}_s \tilde{r}_{c}^3 $ where $ n_{s} $ is the number of salt ions; $\bar{c}_p $ is defined as $ \bar{c}_p=n_{p} /(\Omega / r_c^3)=(n_p \ell_{0}^{3}/\Omega)(r_c^{3}/\ell_{0}^{3})=\tilde{c}_p \tilde{r}_{c}^3 $ where $ n_{p}$ is the number of protons, $c_p \approx [H^+]$, $ \ell_{0} $ is the bond length, and $ \Omega $ is the system volume.

$F_0=F_1+F_2=-k_B T \ln( \mathcal{Z}_{1} \mathcal{Z}_{2})$ (Eqs. \eqref{z1} and \eqref{z2}). The entropic contribution arising from the various distributions of the adsorbed counterions and coions, 
	\begin{align}\label{F1}
		\frac{F_{1}}{k_B T}&=\sum_{i=1}^{3}b_i\log(b_i)-\log{(N_c)},
	\end{align}
    where, $ b_1=(N_c -M_p -M_s),~b_2=M_p,$ and $b_3=M_s$. The translational entropy of the free ions \cite{muthu2004,Kundagrami2008,Muthukumar2012,Ghosh2024}, using 
    \begin{equation}
    \label{F2}
		\frac{F_{2}}{k_B T}=N\left[\sum_{i=1}^{3}(c_i \log(c_i\rho)-c_i)\right],
	\end{equation}
    where, $c_1=(\bar{c}_s/\bar{\rho} -M_s/N),~c_2=(N_{c}/N+\bar{c}_p/\bar{\rho}-M_p/N)$, and $~c_3=(\bar{c}_s+\bar{c}_p)/\bar{\rho}$.

    \mycomment{\begin{align}
    \mathcal{Z}_{2}=\frac{\Omega^{\sum_{\gamma}n_{\gamma}}}{\prod_{\gamma}n_{\gamma}!}
    \end{align}
    where, $n_{\gamma}$ is the free ions, and is defined as the following, $n_{1}=N-M_{p}+n_{H+}$, $n_{2}=n_{OH-}$, $n_{3}=n_{+}-M_{s}$, and $n_{4}=n_{-}$ where $ n_{H+},n_{OH-},n_{-}, \text{and }n_{+} $ are the number of protons, hydroxyl of the protic solvent, anions, and cations of the electrolyte respectively.}

    In the limit of $\kappa \ell_0 \rightarrow 0$, the Helmholtz free energy due to counterion density fluctuations approaches the following expression\cite{muthu2002}:
	\begin{align}
		\frac{F_{3}}{k_{B}T}=-\frac{\Omega \widetilde{\kappa}^3}{12 \pi},
		\label{DH}
	\end{align}
    where $\widetilde{\kappa}$ is the dimensionless inverse Debye length given by $\widetilde{\kappa}=\sqrt{ 4 \pi \widetilde{ \ell}_{B}\sum_{i=1}^{3}(c_i\rho)}$, and $\widetilde{\kappa}=\kappa \ell_{0}$.  Note that the expression given in Eq. \eqref{DH} depends only on the concentration of the dissociated counterions through $\kappa$. However, when $\widetilde{\kappa}>0$, the effects of ionic sizes become relevant. An alternative method, whereby  the limit ($\widetilde{\kappa}>0$) is modified by the finiteness of counterions (taken as spheres of diameter $r_c$) can be derived by adopting the Debye-H\"uckel argument, as described in Ref. \cite{mcquarrie2000}. The resulting equation for the Helmholtz free energy, $F_3$, is given by
	\begin{align}
		\frac{F_{3}}{\bar{\Omega} k_{\mathrm{B}} T}=-\frac{1}{4 \pi}\left[\ln (1+\widetilde{\kappa} \tilde{r}_{c})-\widetilde{\kappa} \tilde{r}_{c}+\frac{1}{2}(\widetilde{\kappa} \tilde{r}_{c})^2\right].\label{F3-mcqurrie}
	\end{align}
	
	This equation, Eq.\eqref{F3-mcqurrie}, reduces to Eq.\eqref{DH} in the salt-free case. In this approach, we pre-integrate the salt degrees of freedom, resulting in a Debye-H\"uckel description of screened interactions between monomer units, which does not influence the effective interaction between the salt ions\cite{muthu2002}. This approximation of using a screened Coulomb interaction for macroions is commonly used in analytical and simulation studies. However, it should be noted that the free energy given by Eqs. \eqref{DH} and \eqref{F3-mcqurrie} differs from those of previous theories, including those presented in Ref. \cite{delacruz2003,delacruz2004}.

    The dimensionless adsorption free energy, following Eq.\eqref{uad}, is
	\begin{equation}
		\frac{F_{4}}{k_{B}T}= - M_s \delta \widetilde{\ell}_B/\widetilde{d} - M_{p}(2.303\,pK_\mathrm{int}),
		\label{F4}
	\end{equation}
	where $ \delta=(\epsilon/\epsilon_{l}) $, and $\widetilde{d}=d/\ell_{0}$ with $d$ being the effective dipole length between the counterion and side chain. Further details are provided in \cref{appendix:binding_constant}.
    The conformational free energy of the poly-acid is
    	\begin{align}
		\frac{F_{pol}}{k_{B}T}&=\frac{3}{2}\left[\widetilde{\ell}_{1}-1-\log \widetilde{\ell}_{1}\right]+ \left(\frac{9}{2 \pi}\right)^{3 / 2}  \frac{w \sqrt{N}}{\widetilde{\ell}_{1}^{3 / 2}} \nonumber\\
		&+\frac{f^2  N^{2}  \widetilde{\ell}_{B}}{2} \Theta_{s}\left(\alpha\right),\label{F5-appendix}
	\end{align}
	where,
	\begin{align}
		\Theta_{s}\left(\alpha\right)=\frac{4}{\pi}\left[\sqrt{\frac{\pi\widetilde{\kappa}^{2}}{4\alpha}}-\frac{\widetilde{\kappa} \pi}{2} \exp{\left(\alpha\right)} \text{erfc}\left(\sqrt{\alpha}\right)\right],
	\end{align}
	$ \alpha={\widetilde{\kappa}^{2} R_{g}^2}/{3} $, and $ \widetilde{ \ell}_{1} $ is the variational parameter that gives the effective expansion factor of the poly-acid compared to its Gaussian size.

We minimize the total free energy Eq.\eqref{partition-sum} or equivalently 
\begin{align}\label{eq:free_energy_sum}
    \mathcal{F}=\sum_{i=1}^4 F_i+F_{pol}, 
\end{align}
using \cref{F1,F2,DH,F4,F5-appendix}, with respect to $M_p,~M_s,$ and $\widetilde{\ell}_1$, such that
\begin{align}\label{eq:free_energy_sum_minimise}
\frac{\partial\mathcal{F}}{\partial M_p}=
\frac{\partial\mathcal{F}}{\partial M_s}=
\frac{\partial\mathcal{F}}{\partial\widetilde{\ell}_1}=0.
\end{align}
The numerical minimization is carried out using a downhill simplex algorithm \cite{Nelder1965}.

\mycomment{\section{Conversion of solution concentrations to dimensionless form}
\label{appendix:conversion_mM}
\subsection{From pH to Proton Concentration}

In dilute aqueous solutions, the pH is operationally defined as the negative base-10 logarithm of the hydrogen ion concentration:
For convenience we have defined
\begin{equation}
    \mathrm{pH} = -\log_{10} [\mathrm{H}^+],
\end{equation}
where $[\mathrm{H}^+]$ is the molar concentration of protons (in mol/L). This definition is strictly valid in the limit of sufficiently dilute solutions, where the system behaves ideally and non-ideal effects arising from electrostatic interactions or ion-specific interactions are negligible.

More generally, pH is defined in terms of the hydrogen ion activity,
\begin{equation}
    \mathrm{pH} = -\log_{10} a_{\mathrm{H}^+},
\end{equation}
with $a_{\mathrm{H}^+} = \gamma_{\mathrm{H}^+}[\mathrm{H}^+]$, where $\gamma_{\mathrm{H}^+}$ is the activity coefficient. However, under dilute conditions ($\gamma_{\mathrm{H}^+} \approx 1$), this reduces to the commonly used concentration-based definition above.

For practical calculations, the proton concentration in millimolar units (mM) is given by
\begin{equation}
    [\mathrm{H}^+]~(\mathrm{mM}) = 10^{-\mathrm{pH}} \times 10^3.
\end{equation}
For example, at $\mathrm{pH} = 7$, $[\mathrm{H}^+] = 10^{-7} \times 10^3 = 0.0001~\mathrm{mM} = 0.1~\mu\mathrm{M}$.
\mycomment{}
\subsection{From mM to Dimensionless Concentration}

To convert any concentration $c$ (in mM) to the dimensionless form $\bar{c}$ as required in the free energy functional, we use
\begin{equation}
    \bar{c} = c~\text{(in mM)} \times \widetilde{r}_c^3 \times 0.0006022\,\ell^3,
\end{equation}
where $\widetilde{r}_c = r_c/\ell$ is the reduced ion (or molecular) radius, and $\ell$ is the characteristic length scale (e.g., Kuhn length). This conversion ensures that all concentrations are consistently represented in the dimensionless units appropriate for the theoretical framework.

\section{Appendix: Model Parameters}\label{appendix:parameters}

Table~\ref{tab:parameters} summarizes the parameters used in the theoretical calculations presented in this work.

\begin{table}[h!]
\centering
\caption{Model parameters used in the numerics}
\begin{tabular}{llc}
\hline
\textbf{Parameter} & \textbf{Description} & \textbf{Value} \\
\hline
$N$           & Number of monomers                 & 52 \\
$N_c$         & Number of acidic monomers         & 44 \\
$w$           & Excluded volume parameter          & 3 \\
$w_3$         & Three-body interaction parameter   & 3 \\
$\delta$      & dielectric mismatch parameter        & 1 \\
$r_c$         & ionic radius                  & 0.34 nm \\
$\ell$        & Kuhn length = $C_\alpha-C_\alpha$ length                    & 0.38 nm \\
$\ell_{B}$    & Bjerrum length                     & 0.72 nm \\
\hline
\end{tabular}
\label{tab:parameters}
\end{table}}

\section{Model Parameters and Numerical Conventions}
\label{appendix:parameters}

For consistency across all theoretical calculations, all concentrations $c$ (in mM) are converted to their dimensionless form $\bar{c}$ using the relation:
\begin{equation}
    \bar{c} = c\,\text{(in mM)} \times \widetilde{r}_c^3 \times 0.0006022\,\ell_0^3,
\end{equation}
where $\widetilde{r}_c = r_c/\ell_0$ is the reduced ionic (or hydration) radius, and $\ell_0$ is the bond length. This conversion ensures that all concentrations are consistently represented in the dimensionless units required by the theoretical framework.

Table~\ref{tab:parameters} summarizes the parameters used in the numerical calculations presented in this work.

\begin{table}[h!]
\centering
\caption{Model parameters used in the numerics.}
\begin{tabular}{llc}
\hline
\textbf{Parameter} & \textbf{Description}         & \textbf{Value} \\
\hline
$N$           & Number of monomers                 & $52$ \\
$N_c$         & Number of acidic monomers           & $44$ \\
$w$           & Excluded volume parameter           & $3$ \\
$w_3$         & Three-body interaction parameter    & $3$ \\
$\delta$      & Dielectric mismatch parameter       & $1$ \\
$\ell_{0}$        & bond length                       & $0.38$ nm \\
$\ell$        & Kuhn length estimated in \cite{Chowdhury2023}                       & $0.8$ nm \\
$r_c$         & Ionic radius                       & $0.38$ nm \\
$\ell_{B}$    & Bjerrum length                     & $0.72$ nm \\
$\rho$  & Monomer density           & $10^{-9}$ \\
\hline
\end{tabular}
\label{tab:parameters}
\end{table}
\vspace{1em}

\section{Semi-flexible Polymer: Edwards–Singh Variational Approach}\label{appendix:semi-flex}
The polymer chain is parameterized by $\mathbf{R}(s)$, with $s \in [0, L]$, and the tangent vector is defined as $\boldsymbol{\tau}(s) = \frac{d\mathbf{R}(s)}{ds}$. Following established approaches \cite{Bawendi1985,Bhattacharjee1987,Lagowski1991,Ghosh2001}, the Hamiltonian for a semi-flexible chain without interactions, is given by
\begin{align}
    H_0 &= \frac{3}{4} \left(|\boldsymbol{\tau}(L)|^2 + |\boldsymbol{\tau}(0)|^2 \right)\nonumber\\
    &+ \frac{3}{2\ell} \int_0^L ds\, |\boldsymbol{\tau}(s)|^2
    + \frac{3\ell}{8} \int_0^L ds\, \left| \frac{d\boldsymbol{\tau}(s)}{ds} \right|^2,\label{eq:semiflex:H_0}
\end{align}
where the three terms enforce global in-extensibility, penalize chain stretching, and account for bending rigidity, respectively. The full Hamiltonian can be written as $H_{semi}=H_0+H_{ex}+H_{el}$, using \cref{eq:semiflex:H_0,eq:H_ex,eq:H_el}.

Following the Edwards-Singh variational procedure\cite{edwards1979,Ghosh2001}, we define a trial Hamiltonian with a renormalized Kuhn length $\ell_1$:
\begin{align}
    H_R &= \frac{3}{4} \left(|\boldsymbol{\tau}(L)|^2 + |\boldsymbol{\tau}(0)|^2 \right)\nonumber\\
    &+ \frac{3}{2\ell_{1}} \int_0^L ds\, |\boldsymbol{\tau}(s)|^2
    + \frac{3\ell_1}{8} \int_0^L ds\, \left| \frac{d\boldsymbol{\tau}(s)}{ds} \right|^2.
\end{align}

The principle behind this is to express observables such as the mean-square end-to-end distance, $\langle R^2 \rangle$, as a perturbative expansion in $H_R - H_{semi}$:
\begin{align}
    \langle R^2 \rangle &= \langle R^2 \rangle_R
    + \langle R^2 \rangle_R \langle H_R - H_{semi} \rangle_R\nonumber\\
    &
    - \langle (H_R - H_{semi}) R^2 \rangle_R+ O\left( (\langle H_R - H_{semi} \rangle_R)^2 \right),
\end{align}
where $\langle \cdot \rangle_R$ indicates an average with respect to the trial Hamiltonian $H_R$. The variational constraint is that $\ell_1$ is chosen such that the mean-square end-to-end distance calculated with $H_R$ matches that of the interacting system, $\langle R^2 \rangle = \langle R^2 \rangle_R,$ which, to leading order, yields the equation
\begin{align}
    \langle R^2 (H_R - H_{semi}) \rangle_R - \langle R^2 \rangle_R \langle H_R - H_{semi} \rangle_R = 0.
\end{align}
The explicit solution for $\ell_1$ can be found in the ref.\cite{Ghosh2001} and references therein.

\end{document}